\documentclass[prb,
showpacs,
twocolumn,
floats,
10pt,
aps,
citeautoscript,
longbibliography,
superscriptaddress]{revtex4-2}

\usepackage{placeins}
\usepackage{lipsum}
\usepackage{xcolor}
\usepackage[normalem]{ulem}
\usepackage{comment}
\usepackage{tabularx}
\usepackage{graphicx}
\usepackage{dcolumn}
\usepackage{bm}
\usepackage{ulem}

\usepackage{bbm}
\usepackage{blindtext}
\usepackage{graphics}
\usepackage{verbatim}   
\usepackage{amsfonts}
\usepackage{amsmath}
\usepackage{amssymb}
\usepackage{adjustbox}
\usepackage{microtype} 
\allowdisplaybreaks 
\usepackage{xspace} 
\usepackage{multirow} 
\usepackage{tabstackengine}
\setstackEOL{\cr}

\usepackage{lmodern}
\usepackage[T1]{fontenc}

\newcommand*{\ndots}{\kern-0.075em.\kern-0.05em.\kern-0.05em.}  
\newcommand*{\nidots}{.\kern-0.05em.\kern-0.05em.} 
\newcommand*{\ncdots}{\kern-0.15em\cdot\kern-0.2em\cdot\kern-0.2em\cdot\kern-0.15em}   

\NewDocumentCommand{\doubleI}{O{}}{\mathbbm{1}_{#1}}
\NewDocumentCommand{\doubleIb}{O{}}{{\overline{\mathbbm{1}}_{#1}}}
\NewDocumentCommand{\doubleIk}{O{}}{\mathbbm{1}^\ks_{\! #1}}
\NewDocumentCommand{\doubleId}{O{}}{\mathbbm{1}^\ds_{\! #1}}
\NewDocumentCommand{\doubleIp}{O{}}{\mathbbm{1}^\ps_{\! #1}}
\NewDocumentCommand{\doubleV}{O{}}{\mathbbm{V}_{\! #1}}
\NewDocumentCommand{\doubleVk}{O{}}{\mathbbm{V}^\ks_{\! #1}}
\NewDocumentCommand{\doubleVd}{O{}}{\mathbbm{V}^\ds_{\! #1}}
\NewDocumentCommand{\doubleVp}{O{}}{\mathbbm{V}^\ps_{\! #1}}
\NewDocumentCommand{\doublev}{o}{{\mathbbm{v}_{#1}}}
\NewDocumentCommand{\doubleVb}{o}{{\overline{\mathbbm{V}}_{\! #1}}}
\NewDocumentCommand{\doubleVt}{o}{{\widetilde{\mathbbm{V}}_{\! #1}}}
\NewDocumentCommand{\doubleVh}{o}{\widehat{{\mathbbm{V}}_{\! #1}}}
\NewDocumentCommand{\doubleW}{o}{\mathbbm{W}_{\! #1}}
\NewDocumentCommand{\doubleWk}{o}{\mathbbm{W}^\ks_{\! #1}}
\NewDocumentCommand{\doubleWd}{o}{\mathbbm{W}^\ds_{\! #1}}
\NewDocumentCommand{\doubleWb}{o}{{\overline{\mathbbm{W}}_{\! #1}}}
\NewDocumentCommand{\doubleWt}{o}{{\widetilde{\mathbbm{V}}_{\! #1}}}
\NewDocumentCommand{\doubleWh}{o}{{\widehat{\mathbbm{V}}_{\! #1}}}

\newcommand{\LMUMunich}{Department of Physics and Arnold Sommerfeld Center for Theoretical Physics (ASC), Ludwig-Maximilians-University Munich,
Theresienstr. 37, D-80333 Munich, Germany}

\definecolor{darkgreen}{rgb}{0,0.5,0}
\definecolor{purple}{rgb}{0.6,0,0.5}
\definecolor{orange}{rgb}{1,0.5,0}
\definecolor{darkred}{rgb}{.7,0,0}
\definecolor{darkblue}{rgb}{0,0,.6}
\definecolor{grey}{rgb}{.6,.6,.6}
\definecolor{dimgreen}{rgb}{0.2,0.7,0.2}
\definecolor{brightgreen}{rgb}{0.5020, 1, 0}

\newcommand{\jvdomit}[1]{}

\usepackage{multirow}
\usepackage{physics}
\usepackage{hyperref}
\hypersetup{colorlinks=true,breaklinks,linkcolor=blue,urlcolor=blue,citecolor=blue}

\usepackage{graphicx}
\usepackage{dcolumn}
\usepackage{bm}

\usepackage[caption=false]{subfig}
\usepackage{braket}

\begin{document}

\preprint{}

\title{Improved contraction of finite projected entangled pair states}


\author{Markus Scheb}%
\affiliation{\LMUMunich}





\date{\today}

\begin{abstract}
We present an improved version of the algorithm contracting and optimizing finite projected 
entangled pair states (fPEPS) in conjunction with projected entangled pair operators (PEPOs). Our work
has two
components to it. First, we explain in detail the characteristic contraction patterns that occur in
fPEPS calculations and how to slice them such that peak memory occupation remains
minimal while ensuring efficient parallel computation. Second, we combine controlled bond expansion
[A. Gleis, J.-W. Li, and J. von Delft, 
\href{https://journals.aps.org/prl/abstract/10.1103/PhysRevLett.130.246402}{Phys. Rev. Lett. \textbf{130}, 246402 (2023)}]
with randomized singular value decomposition 
[V. Rokhlin, A. Szlam, and M. Tygert, \href{https://epubs.siam.org/doi/abs/10.1137/080736417}{SIAM J. Matrix Anal. Appl. (2009)}]
and apply
it throughout the fPEPS algorithm. We present benchmark results for the Hubbard model for system
sizes up to 8$\times$8 and SU(2) symmetric bond dimension of up to $D=6$ for PEPS bonds and $\chi=500$
for the environment bonds. Finally, we comment on the state and future of the fPEPS-PEPO framework.

\end{abstract}


\maketitle

\section{Introduction}
\label{sec:introduction}
In the past 30 years, tensor networks have become an increasingly popular tool for
calculating quantum many-body systems \cite{Orus2019Sep,Cirac2021Dec,Banuls2023Mar}. The prototype
of this family of algorithms is the density matrix renormalization group (DMRG)
\cite{White1992Nov,White1993Oct}, which operates on a type of one-dimensional 
tensor network state called
matrix product state (MPS) \cite{Rommer1997Jan,Schollwock2011Jan}. MPSs are constructed by 
factorizing and truncating a many-body wavefunction and are most suited for one-dimensional 
quantum systems. Their natural generalization to two dimensions are projected entangled-pair states
(PEPS) \cite{Verstraete2004Jul,Perez-Garcia2007Jul,Schuch2010Oct,Verstraete2006Jun}, 
whose arrangement of tensors matches that of a $2$d lattice. 
While MPSs are convenient to process due to the existence of a canonical gauge, working with a PEPS
has proven to be much more complicated due to its loops and the resulting high costs of tensor 
contractions, the inability to compute expectation values exactly and the poor convergence of 
variational calculations.

The most prominent variant of PEPS-algorithms is the infinite PEPS (iPEPS) 
algorithm \cite{Jordan2008Dec}. It has been successfully applied to various toy models in two 
dimensions \cite{Corboz2016Jul,Phien2015Jul,Corboz2011Jul,Corboz2016Jan,Zhang2025Mar,Li2021Feb}, but
is limited to small unit cells. In addition, PEPSs have been used in more special applications.
For instance, imposing unitarity along all virtual bonds yields isometric tensor networks 
\cite{Zaletel2020Jan,Soejima2020Feb,Lin2022Dec,Kadow2023May}, which are easier to process but also 
limited to more exotic phases, such as string-net liquids. Gaussian PEPSs have been used 
as the starting point for calculating lattice gauge theories \cite{Emonts2023Jul}, d-wave 
superconductors \cite{Yang2023Mar} and U(1)-Dirac spin liquid states \cite{Li2023Feb}. 
Furthermore, PEPSs have been combined with Monte Carlo methods \cite{Vieijra2021Dec}, were shown to 
represent chiral spin liquids \cite{Hasik2022Oct} and were applied to thermal states 
\cite{Sinha2024Jan}.

In this work, we return to the original idea of calculating a grid of unbiased tensors for open 
boundary conditions without any gauge constraints, called finite PEPS (fPEPS) 
\cite{Lubasch2014Aug,Lubasch2014Mar,Scheb2023Apr}. While the methods listed above are restricted 
to either small unit cells or limiting gauges and therefore special phases, fPEPSs without gauge 
constraints have at least the theoretical possibility to describe the entire physical behaviour of 
large, heterogeneous two-dimensional quantum systems in an unbiased fashion.
Sec.~\ref{sec:fpeps} gives a short overview of
the fPEPS-PEPO methodology that was detailed in the precursor of this paper \cite{Scheb2023Apr}.
In Sec.~\ref{sec:optcont}, we describe in detail how to optimally contract the two dominant tensor
clusters that occur while optimizing fPEPSs. Afterwards, we combine the controlled bond expansion 
\cite{Gleis2023Jun} with the randomized singular value decomposition \cite{Rokhlin2009Aug,Halko2011May}
in Sec.~\ref{sec:cbersvd} and apply it to both the environment approximation, as well as the energy 
minimization within the fPEPS algorithm. Afterwards, we present benchmark results for the Hubbard model
in Sec.~\ref{sec:results} and comment on the improvements over the previous version of the algorithm.
Finally, in Sec.~\ref{sec:SummaryOutlook}, we comment on the state and future of the fPEPS-PEPO scheme.

\section{The fPEPS framework}
\label{sec:fpeps}
\begin{figure}[ht]
\includegraphics[width = 0.4\textwidth]{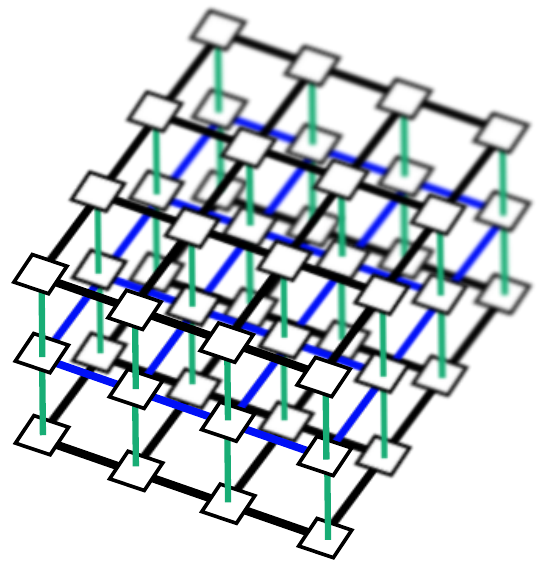}
\label{fig:fpeps}
\caption{PEPS-PEPO network for a 4$\times$4 lattice. Black lines connect PEPS-tensors, blue lines connect
PEPO tensors and green lines connect PEPS and PEPO tensors.}
\end{figure}

In the following, we briefly sketch how to conduct energy minimization through finite PEPSs. 
Fig.~\hyperref[fig:fpeps]{1} illustrates the energy functional 
$E=\left<\psi \right| H \left| \psi \right>$ as a sandwich of a PEPS $\left|\psi\right>$, PEPO 
$H$ and adjoint PEPS $\left<\psi\right|$ for a 4$\times$4 lattice.
The PEPS is a representation of the wavefunction for a two-dimensional system, where two adjacent
tensors are connected by a black bond of dimension $D$. The PEPO stands for a local Hamiltonian 
and is assigned via finite state machines with a blue bond of dimension $w$. The three layers are
connected by green bonds representing local Hilbert spaces of dimension $d$.

Due to the loops inherent to an fPEPS-network, the costs of computing expectation values exactly scale
exponentially with system size. Therefore, a feasible way of working with fPEPSs includes an 
environment approximation, in which bundles of three bonds of total dimension $D\,w\,D$ are
successively compressed to a cumulative
bond of dimension $\chi$. In practical simulations, $\chi \gg D,w,d$.

Given this setup, energy minimization is performed by choosing one of two optimization schemes. 
In the first, called local update, one sweeps over the lattice in both directions to 
optimize a single tensor plus an adjacent bond. The second, called gradient update, allows one
to optimize all PEPS-tensors simultaneously while keeping the basis along the bonds fixed. 

A thorough explanation of the procedure listed above is given in Ref.~\cite{Scheb2023Apr}.

\section{Optimal contraction sequence}
\label{sec:optcont}
\begin{figure}[ht]
\begin{minipage}[c]{\linewidth}
\includegraphics[width = \textwidth]{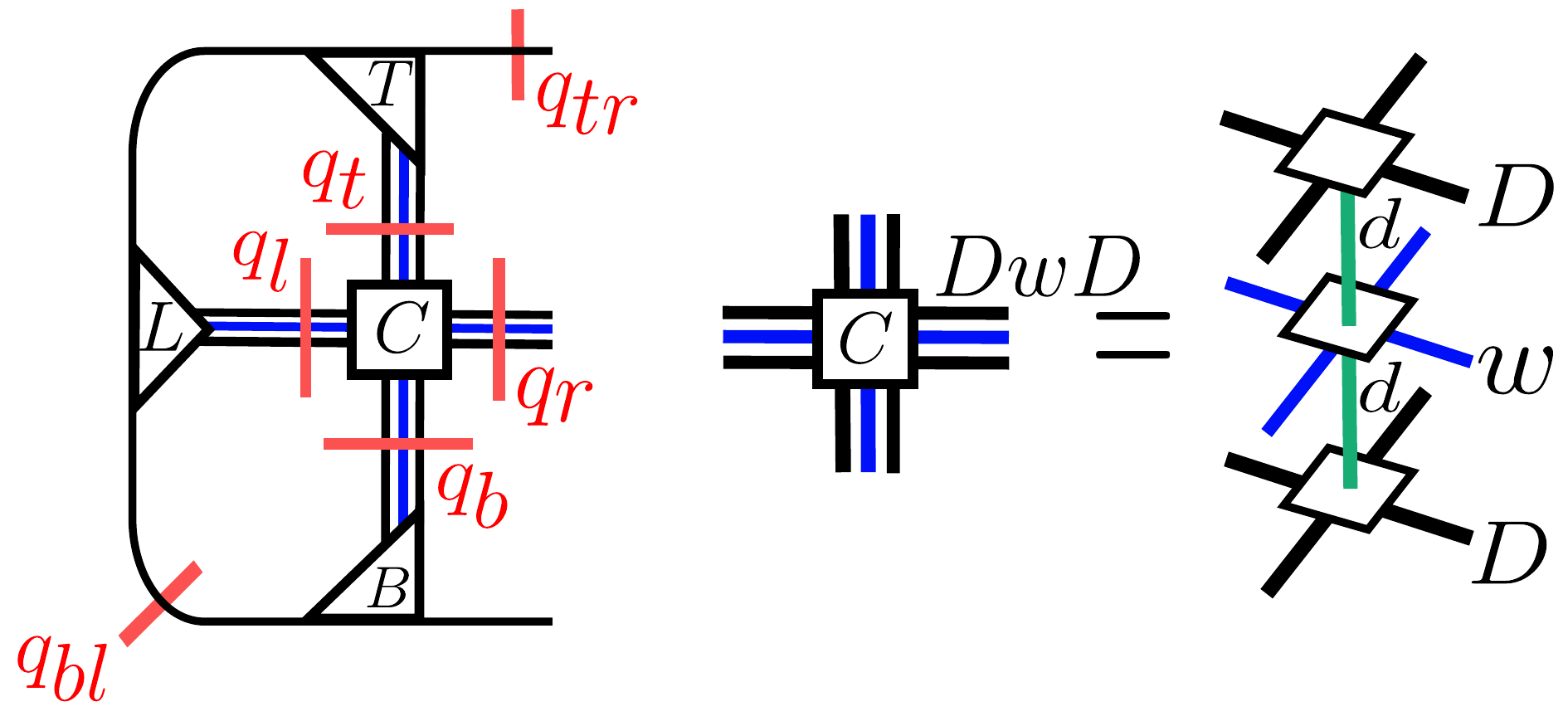}
\centerline{(a)}
\end{minipage}
\\
\begin{minipage}[c]{\linewidth}
\includegraphics[width = \textwidth]{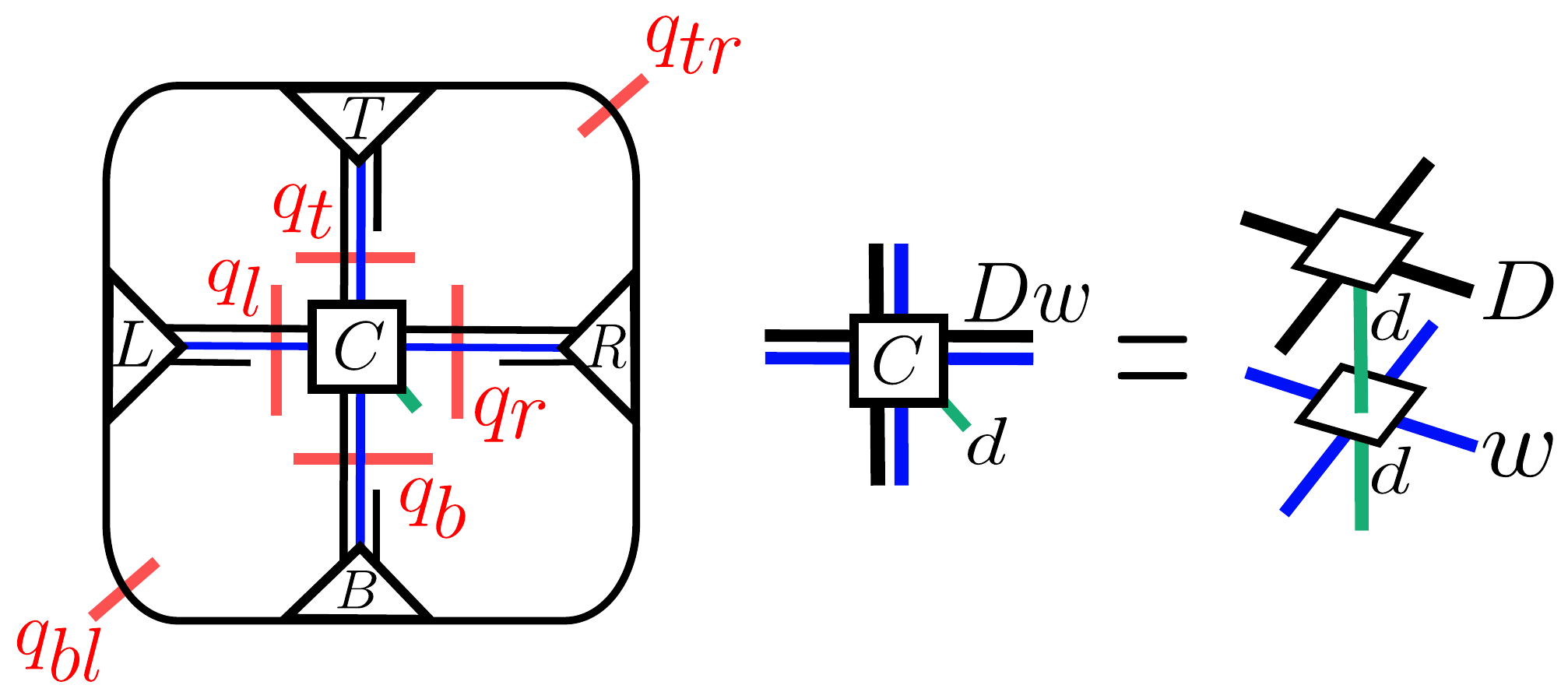}
\centerline{(b)}
\end{minipage}
\label{fig:cont_pattern}
\caption{
Characteristic contraction patterns appearing in the fPEPS 
algorithm, with the environment contraction depicted in (a) and the
contraction of an effective single-site Hamiltonian and a PEPS tensor 
depicted in (b). $T$, $L$, $B$ and $R$ are environment tensors, while the
tensor in the center $C$ is a sandwich of a PEPS tensor, PEPO tensor 
and adjoint PEPS tensor for (a) and a sandwich of a PEPS tensor and 
PEPO tensor for (b).}
\end{figure}

The costs of the fPEPS algorithm are dominated by two characteristic contraction patterns, depicted in
Fig.~\hyperref[fig:cont_pattern]{2}.
\raggedbottom
Fig.~\hyperref[fig:cont_pattern]{2(a)} consists of three environment tensors ($T$,$L$,$B$) plus a 
PEPS-PEPO-PEPS-sandwich ($C$), and is computed during the environment approximation, as well as 
the sweeping 
process at energy minimization. Fig.~\hyperref[fig:cont_pattern]{2(b)} consists of four evironment tensors 
($T$,$L$,$B$,$R$) plus a PEPS-PEPO-sandwich ($C$), and constitutes the 
$H_{\text{eff}} \left| \psi \right>$ operation 
during the Davidson algorithm, where $H_{\text{eff}}$ is the effective Hamiltonian of the single-site 
Hilbert space and $\left| \psi \right>$ is a PEPS-tensor. By removing $T$ and $B$ and the bonds 
attached, one gets the corresponding operations of the DMRG.

Multiplying one tensor after another and as a whole generates giant intermediate contraction results
of size $\order{\chi^2 (DwD)^2}$, which can exceed the size of all other tensors stored in memory.
Therefore, we have developed a strategy for slicing those tensor 
contractions such that peak memory usage remains
as small as possible, without any loss of speed. First, we compute and 
store the sandwich tensor $C$. Since PEPO
tensors of local Hamiltonians exhibit structure beyond mere quantum 
number conservation, we advise
against fusing the indices. Second, we scan $C$ once and associate 
bundles of quantum numbers at the top and 
left ($q_t,q_l$) with their counterparts at the bottom and right 
($q_b,q_r$). This way, 
we generate a map $\left\{(q_t,q_l)\right\}_i \rightarrow 
\left\{(q_b,q_r)\right\}_i$, where the index $i$ designates a set of 
different quantum numbers whose contraction results may add up to the 
same final tensor, should $q_{tr}$ and $q_{bl}$ also be equal. The number 
of these sets determines the number of iterations in the outermost loop 
of our contraction scheme. Third, 
we generate three maps 
(($q_{tr},q_t) \rightarrow \left\{T\right\}_{q_{tr},q_t}$),
(($q_{bl}, q_l) \rightarrow \left\{L\right\}_{q_{bl},q_l}$),
(($q_{bl}, q_b) \rightarrow \left\{B\right\}_{q_{bl},q_b}$),
plus a fourth map 
(($q_{tr}, q_r) \rightarrow \left\{R\right\}_{q_{tr},q_r}$) 
for the $H_{\text{eff}} \left| \psi \right>$ contraction in Fig.~\hyperref[fig:cont_pattern]{2(b)}. These additional maps associate 
the external quantum numbers to the dense blocks inside the environment 
tensors. After these
preparations, we actually calculate the contraction result by nested looping over $i$, $q_{tr}$, 
$q_{bl}$ and
computing $((T \cdot L) \cdot C) \cdot B$ for 
Fig.~\hyperref[fig:cont_pattern]{2(a)} and 
$((T \cdot L) \cdot C) \cdot (B \cdot R)$ for
Fig.~\hyperref[fig:cont_pattern]{2(b)}. Since 
contractions for different $\left(i, q_{tr}, q_{bl}\right)$ do not 
overlap with each 
other, this scheme is easily parallelizable.

\section{Controlled bond expansion via randomized singular value decomposition}
\label{sec:cbersvd}
Processing fPEPSs requires two stages at which a bond between two adjacent tensors is 
optimized. One takes place during the energy minimization of the wavefunction (as is also done in DMRG), the other occurs during the approximation of each environment and is 
structurally identical to an MPS-compression. To circumvent the costs associated with a 
straightforward $2s$-type algorithm, the controlled bond expansion (CBE) \cite{Gleis2023Jun}
allows one to optimize a bond at $1s$ cost.

\begin{figure}[ht]
\includegraphics[width = 0.2\textwidth]{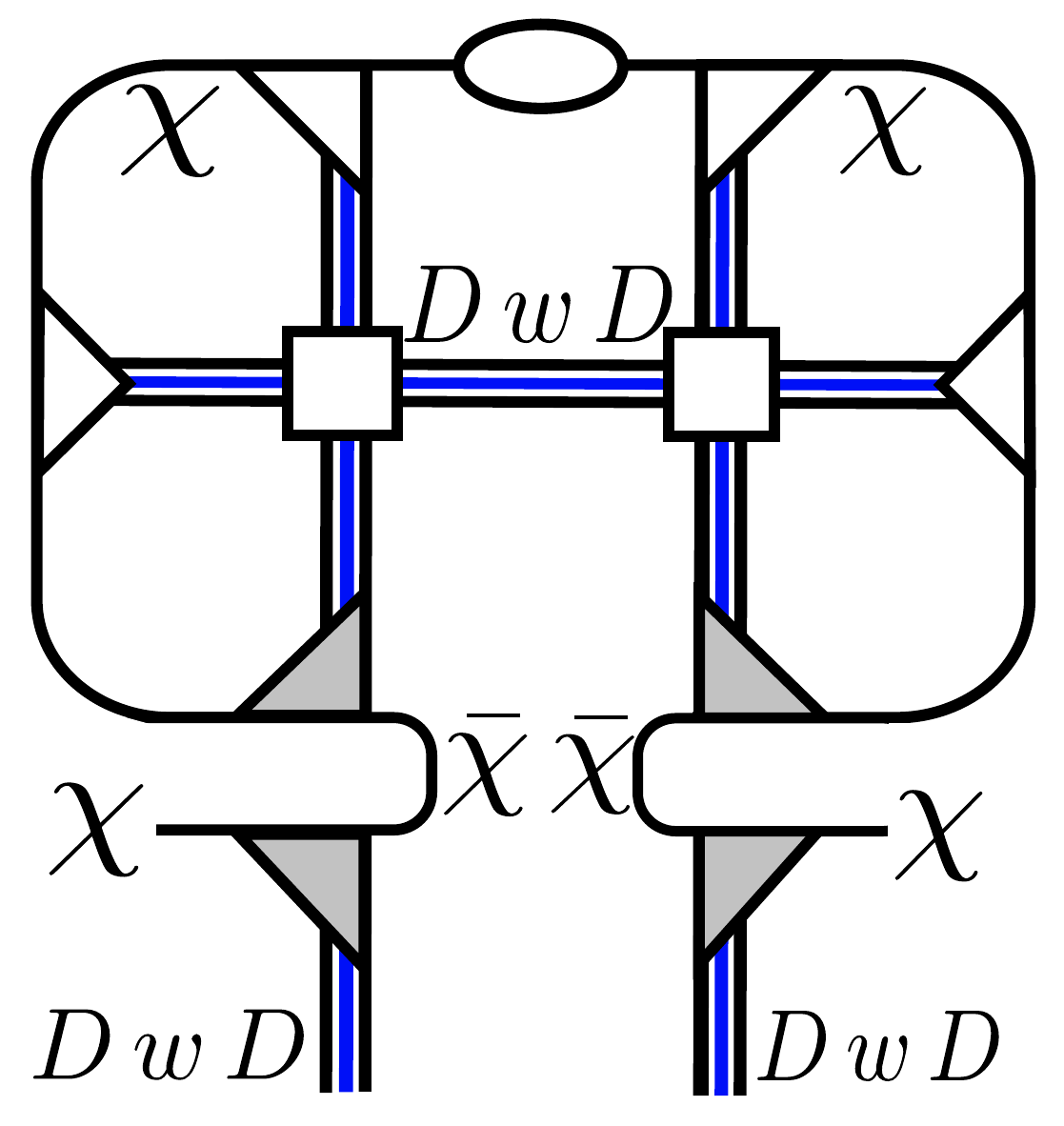}
\label{fig:env_cbe}
\caption{Controlled bond expansion for the environment approximation.}
\end{figure}

\begin{figure}[ht]
\includegraphics[width = 0.3\textwidth]{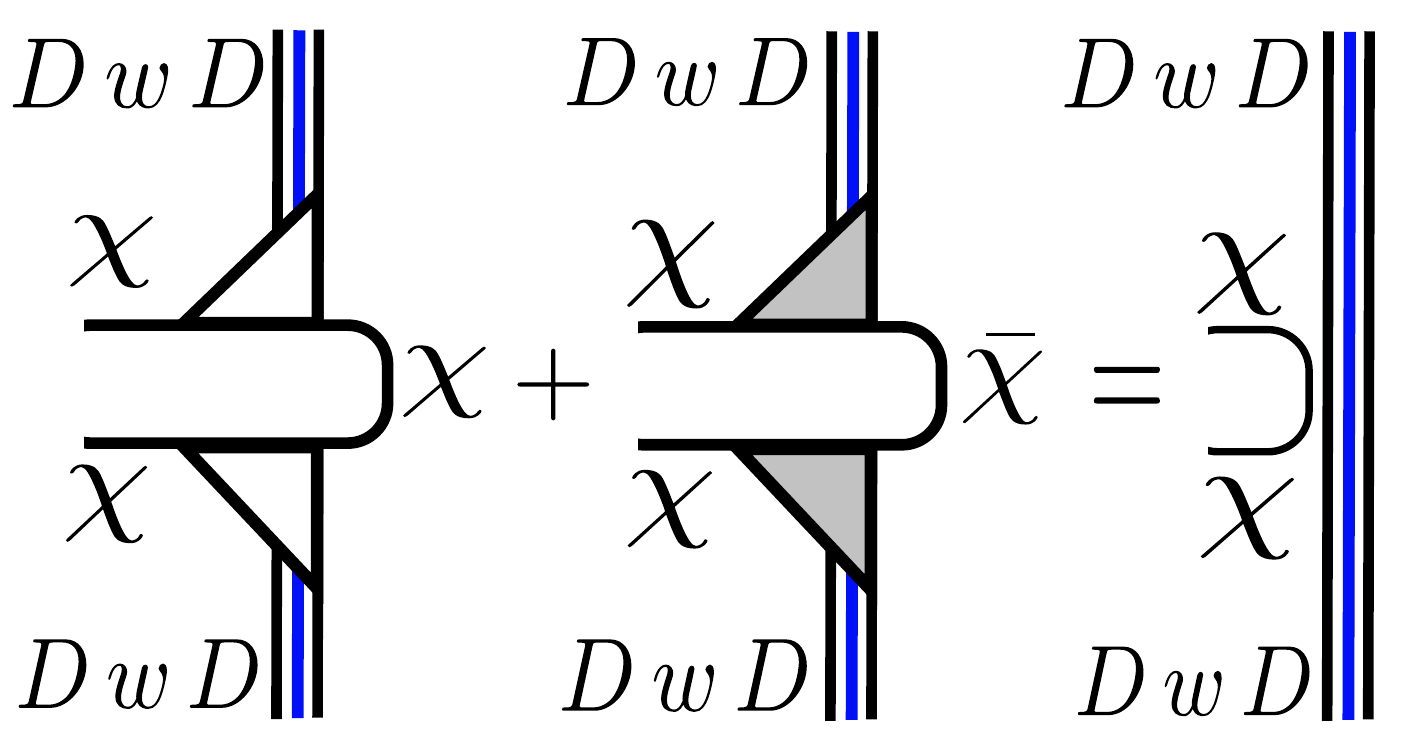}
\label{fig:env_completeness}
\caption{Completeness relation for environment tensors.}
\end{figure}

Fig.~\hyperref[fig:env_cbe]{3} illustrates the CBE for the environment approximation.
The upper row constitutes the previous environment in a mixed canonical form with bond
dimension $\chi$. The middle row is a sequence of PEPS-PEPO-PEPS sandwiches with $D$ as
the PEPS-dimension and $w$ as the PEPO-dimension. The lower row is the new environment 
to be calculated and is supposed to have maximum overlap with the two rows above at 
bond dimension $\chi$. The orthogonal projectors at the bottom are defined by the 
completeness relation in Fig.~\hyperref[fig:env_completeness]{4}.
In the language of projector formalism, $\chi$ and $\bar{\chi}$ are the dimensions of 
kept and discarded space, respectively \cite{Gleis2022Nov}.

Contracting and factorizing the entire cluster in Fig.~\hyperref[fig:env_cbe]{3} generates 
the truncated complement, which contains the most weighty states of the discarded space and
is the final output of the CBE. Processing Fig.~\hyperref[fig:env_cbe]{3} in
this straightforward manner requires operations that are as expensive as performing $2s$ 
optimizations, which is why the CBE was introduced in conjunction with the shrewd selection 
\cite{Gleis2023Jun}, a sequence of contractions and factorizations of smaller tensors. 
However, as was pointed out by McCulloch et al.~\cite{McCulloch2024Mar}, a more efficient 
factorization of a large matrix of small rank can be performed using randomized singular value 
decomposition (RSVD) \cite{Rokhlin2009Aug,Halko2011May}. For the 
$(\chi D w D)\times(\chi D w D)$-matrix $A$ in Fig.~\hyperref[fig:env_cbe]{3}, this scheme 
starts by generating a $(\chi D w D)\times (\tilde{\chi})$-matrix $\Omega$ filled with 
Gaussian random numbers, where $\tilde{\chi} \ll \chi$ is the number of states one wishes 
to extract from the discarded space. Through repeated application of $A$ and $A^T$ onto $\Omega$, 
one can extract the dominant subspace within $A$ and perform an optimized, truncated 
factorization. In the context of CBE for fPEPS, we found a single application of $A$ to 
$\Omega$ to be sufficient.

\begin{figure*}[ht]
\begin{minipage}[c]{0.2\linewidth}
\includegraphics[width = \textwidth]{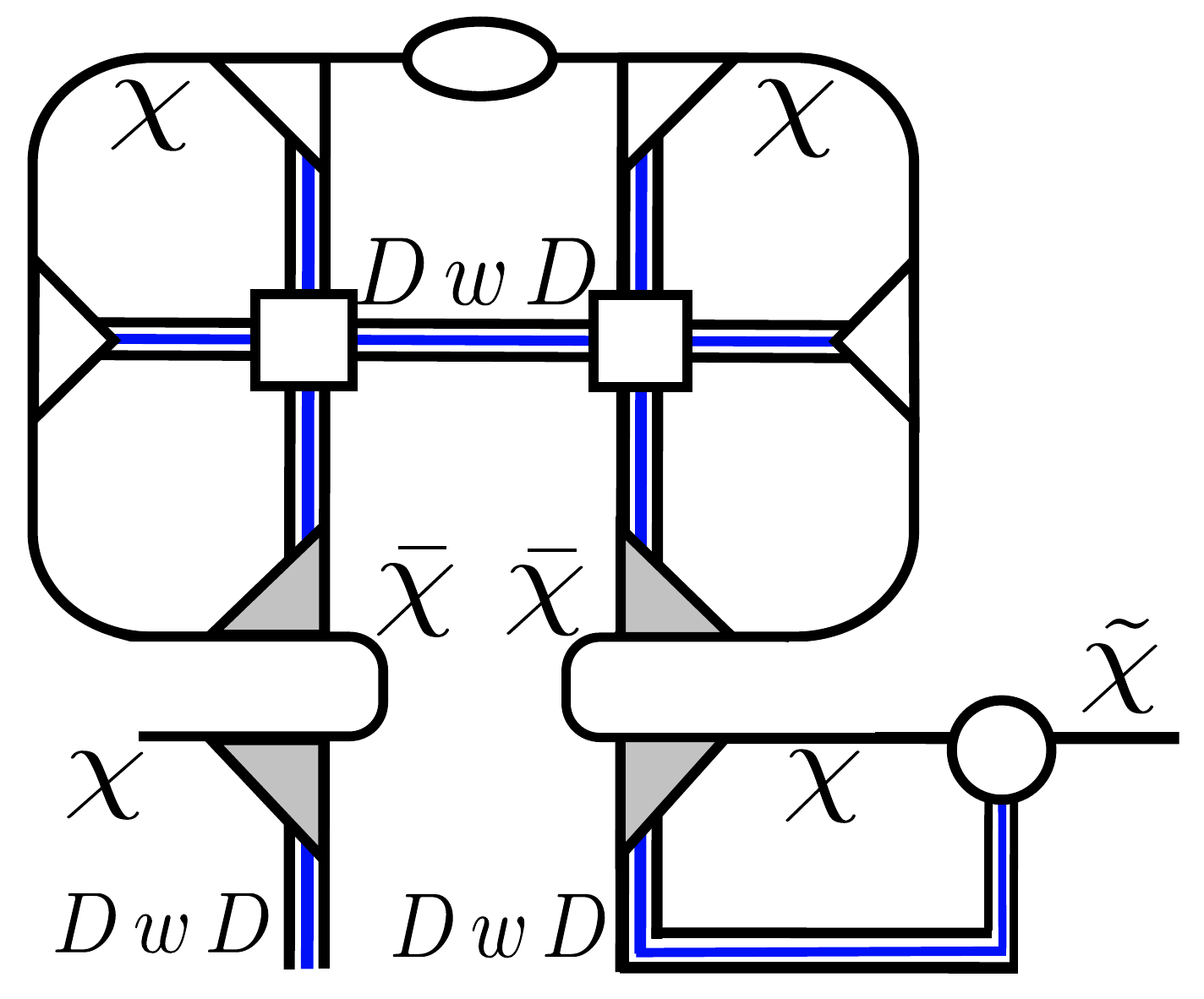}
\centerline{(a)}
\vspace{0.1cm}
\end{minipage}
\hfill
\begin{minipage}[c]{0.35\linewidth}
\includegraphics[width = \textwidth]{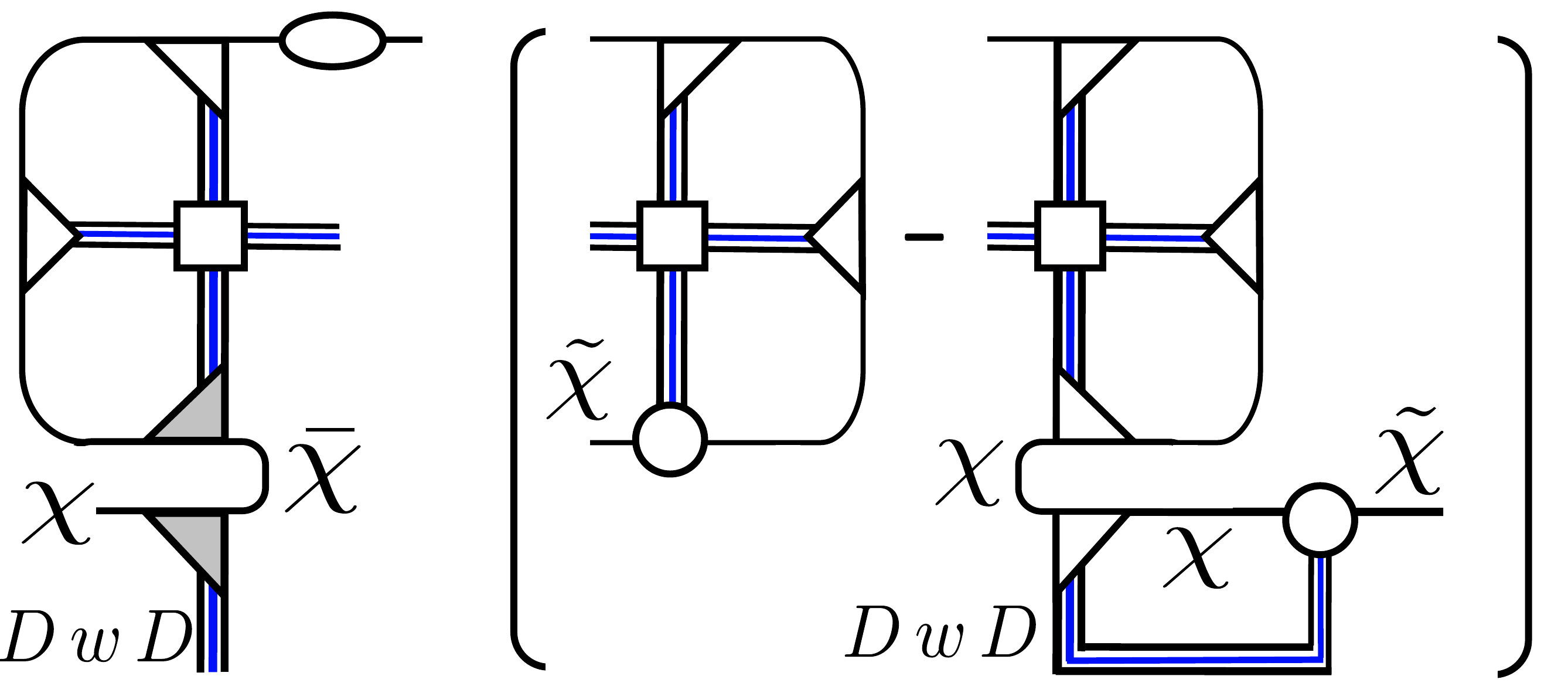}
\centerline{(b)}
\vspace{0.1cm}
\end{minipage}
\hfill
\begin{minipage}[c]{0.35\linewidth}
\includegraphics[width = \textwidth]{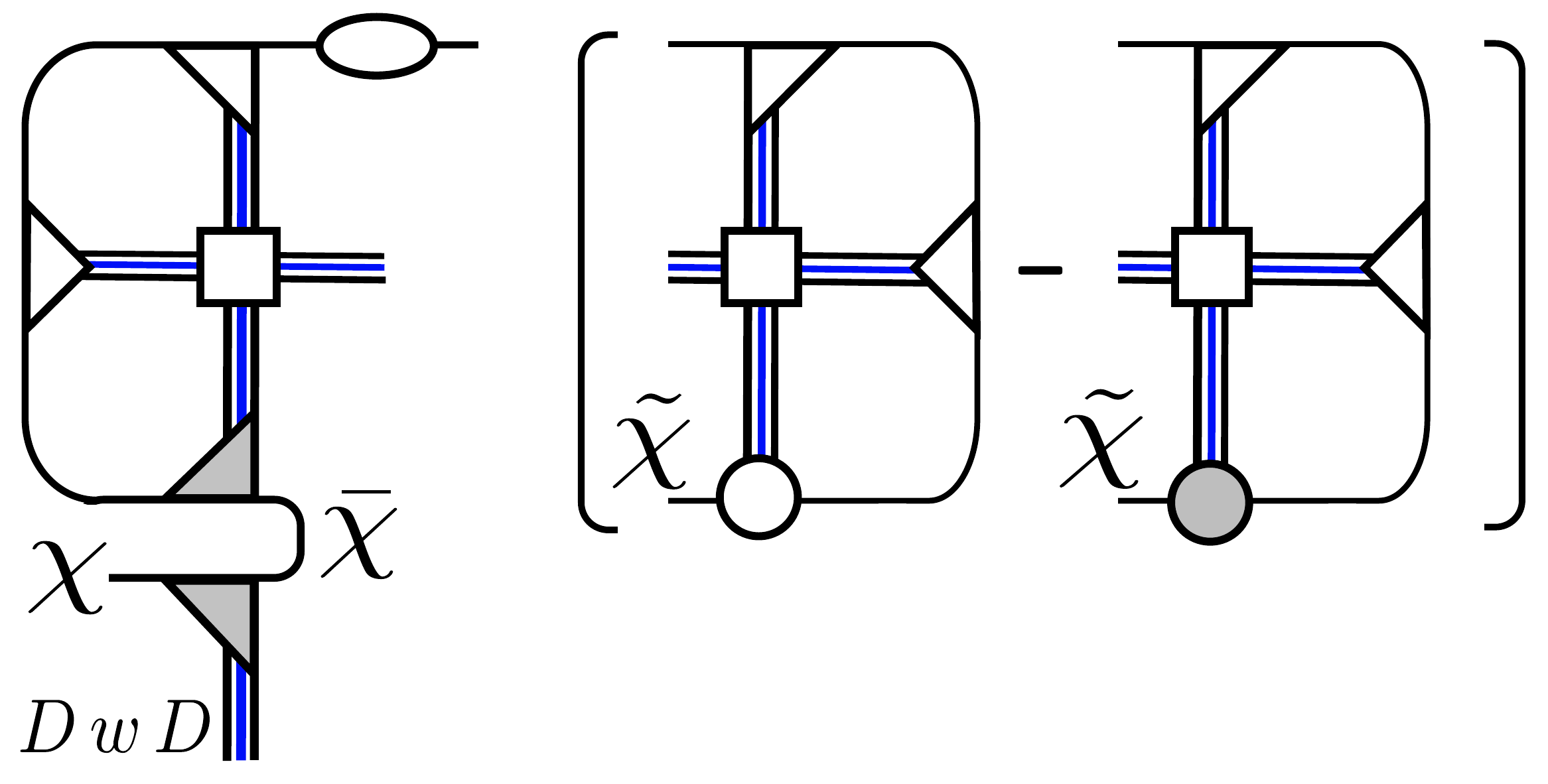}
\centerline{(c)}
\vspace{0.1cm}
\end{minipage}
\\
\begin{minipage}[c]{0.3\linewidth}
\includegraphics[width = \textwidth]{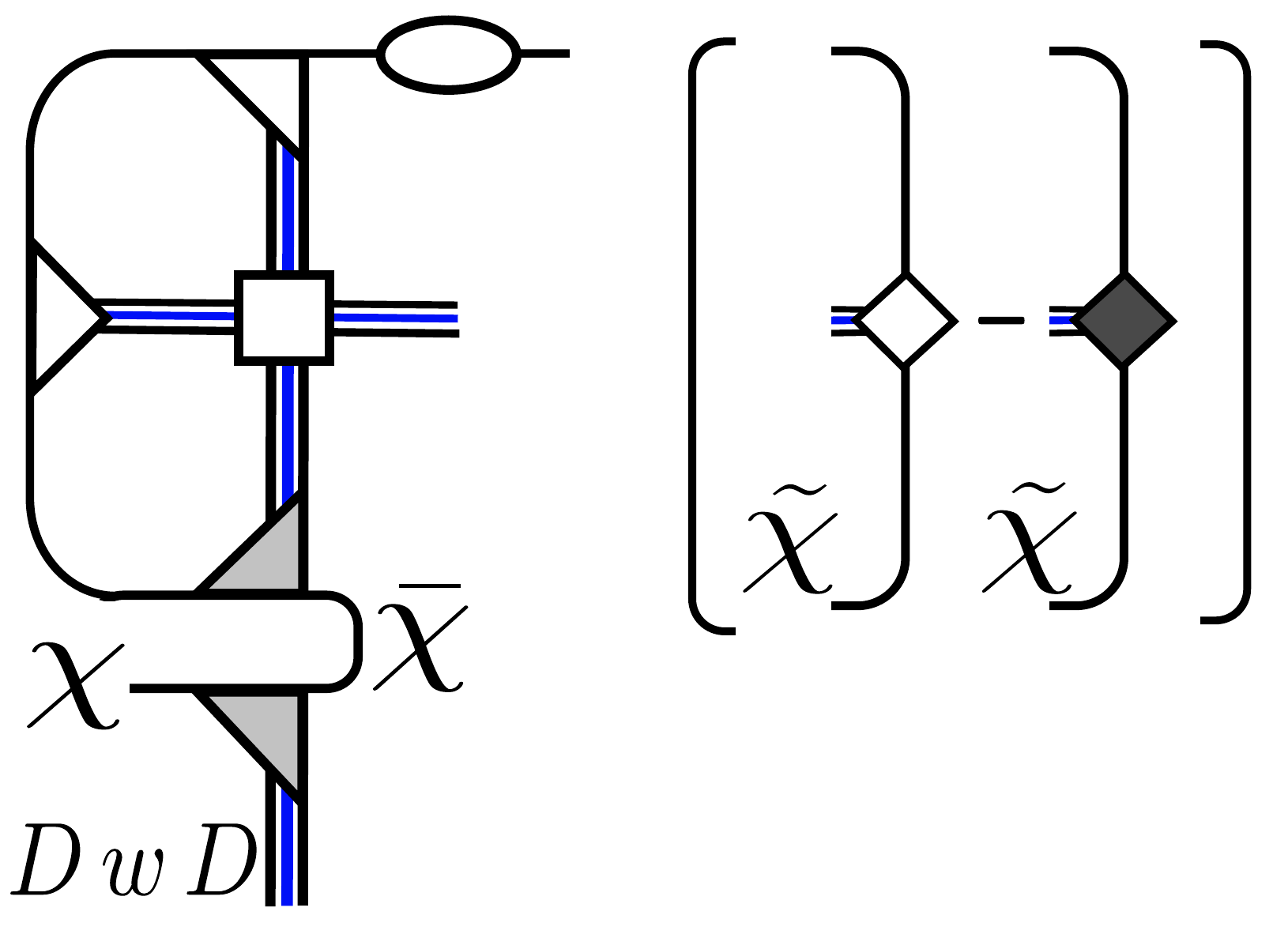}
\centerline{(d)}
\vspace{0.1cm}
\end{minipage}
\hfill
\begin{minipage}[c]{0.15\linewidth}
\includegraphics[width = \textwidth]{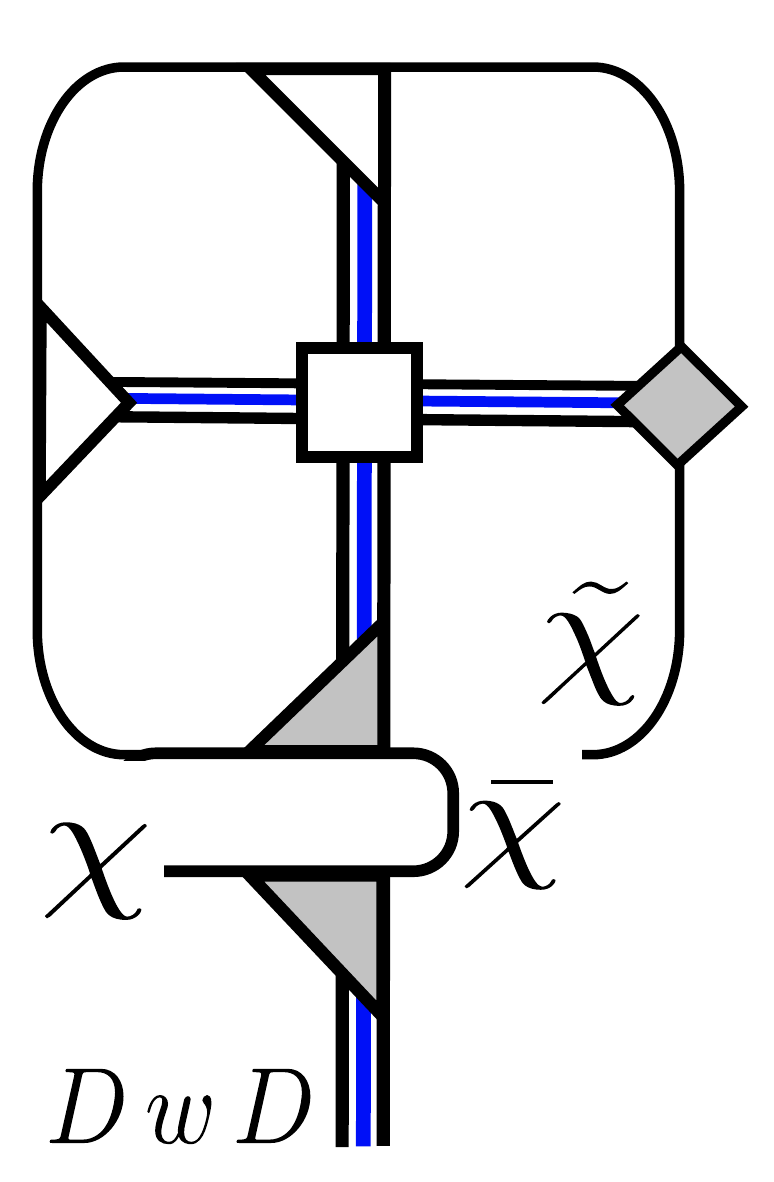}
\centerline{(e)}
\vspace{0.1cm}
\end{minipage}
\hfill
\begin{minipage}[c]{0.2\linewidth}
\includegraphics[width = \textwidth]{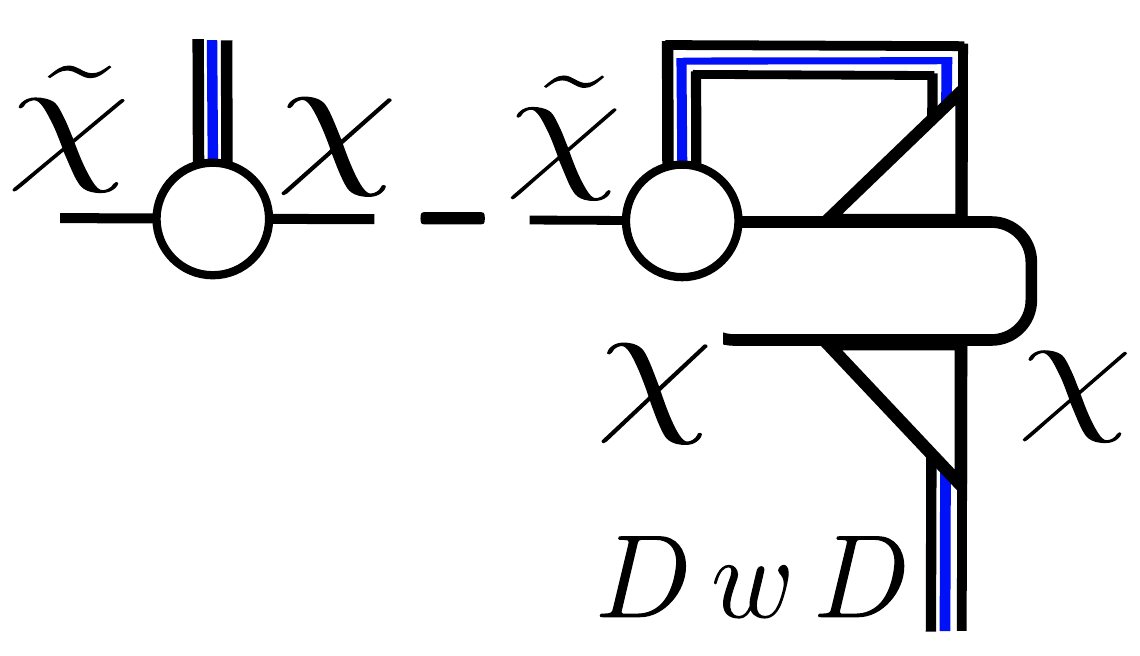}
\centerline{(f)}
\vspace{0.1cm}
\end{minipage}
\hfill
\begin{minipage}[c]{0.1\linewidth}
\includegraphics[width = \textwidth]{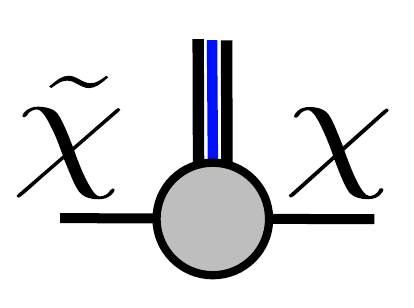}
\centerline{(g)}
\vspace{0.1cm}
\end{minipage}
\label{fig:env_cbe_steps}
\caption{RSVD for the CBE within environment approximation (a) Initial setup of $A \, \Omega$. (b) Separation of right orthogonal projector
into identity and tangential projector via the completeness relation
in Fig.~\hyperref[fig:env_completeness]{4}. (c) Contraction of 
tangential projector and $\Omega$. (d) Contraction of both bracketed clusters. (e) Subtraction
of both bracketed contraction results. (f) Contraction of upper four tensors from (e) and separation
of left orthogonal projector into identity and tangential projector. (g) Final contractions
and subtraction.}
\end{figure*}

For all steps of the RSVD, we refer to Example $1.6$ in 
Ref.~\cite{Halko2011May}. Here, we only detail the 
individual operations of $A \, \Omega$ in Fig.~\hyperref[fig:env_cbe_steps]{5},
which are devised such that the most expensive contraction has 
$\order{\tilde{\chi} \, \chi^2 (DwD)^2}$ cost.
Fig.~\hyperref[fig:env_cbe_steps]{5(a)} illustrates the initial setup with the Gaussian matrix $\Omega$ 
as a white circle on the right. First, we note that one should not calculate the orthogonal 
projector explicitly to avoid generating tensor-legs with a dimension of $\bar{\chi}$. Therefore, 
it is split into the identity and the tangential projector in Fig.~\hyperref[fig:env_cbe_steps]{5(b)}. 
Afterwards, $\Omega$ is contracted with its adjacent environment tensor, leading to two structurally
identical clusters in Fig.~\hyperref[fig:env_cbe_steps]{5(c)}. Both are processed according to 
Fig.~\hyperref[fig:cont_pattern]{2(a)} and subtracted afterwards (Fig.~\hyperref[fig:env_cbe_steps]
{5(d)}). 
In 
Fig.~\hyperref[fig:env_cbe_steps]{5(e)}, the
four tensors on top are again contracted according to Fig.~\hyperref[fig:cont_pattern]{2(a)}, leaving only 
a trivial contraction and subtraction as shown in Fig.~\hyperref[fig:env_cbe_steps]{5(f)} and 
Fig.~\hyperref[fig:env_cbe_steps]{5(g)}.
Note that this new approach to the CBE renders the
operations in Fig.~$19$ and Fig.~$20$ of Ref.~\cite{Scheb2023Apr} obsolete.

\begin{figure}[ht]
\includegraphics[width = 0.25\textwidth]{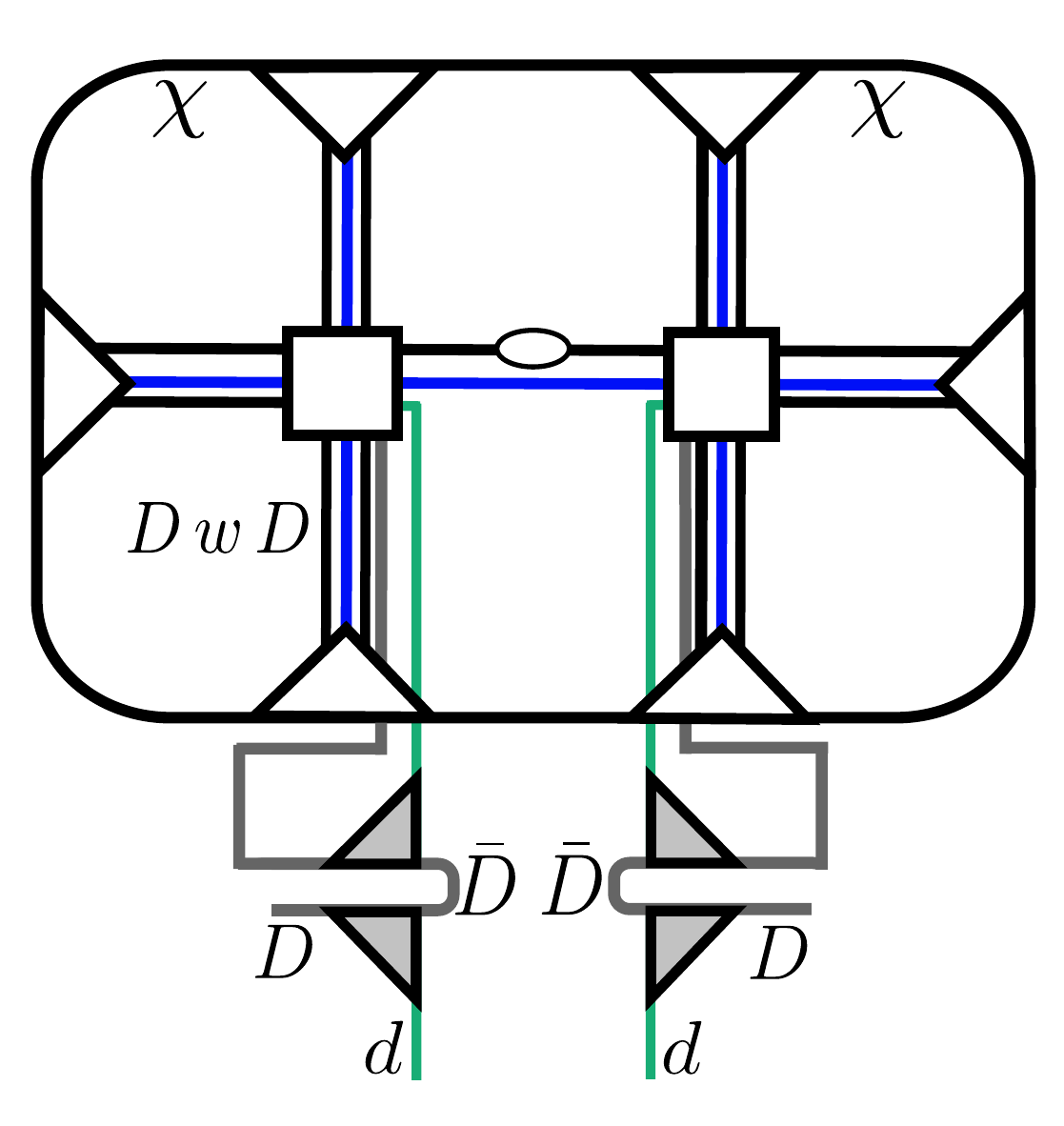}
\label{fig:peps_cbe}
\caption{Controlled bond expansion for energy minimization.}
\end{figure}

\begin{figure*}[ht]
\begin{minipage}[c]{0.2\linewidth}
\includegraphics[width = \textwidth]{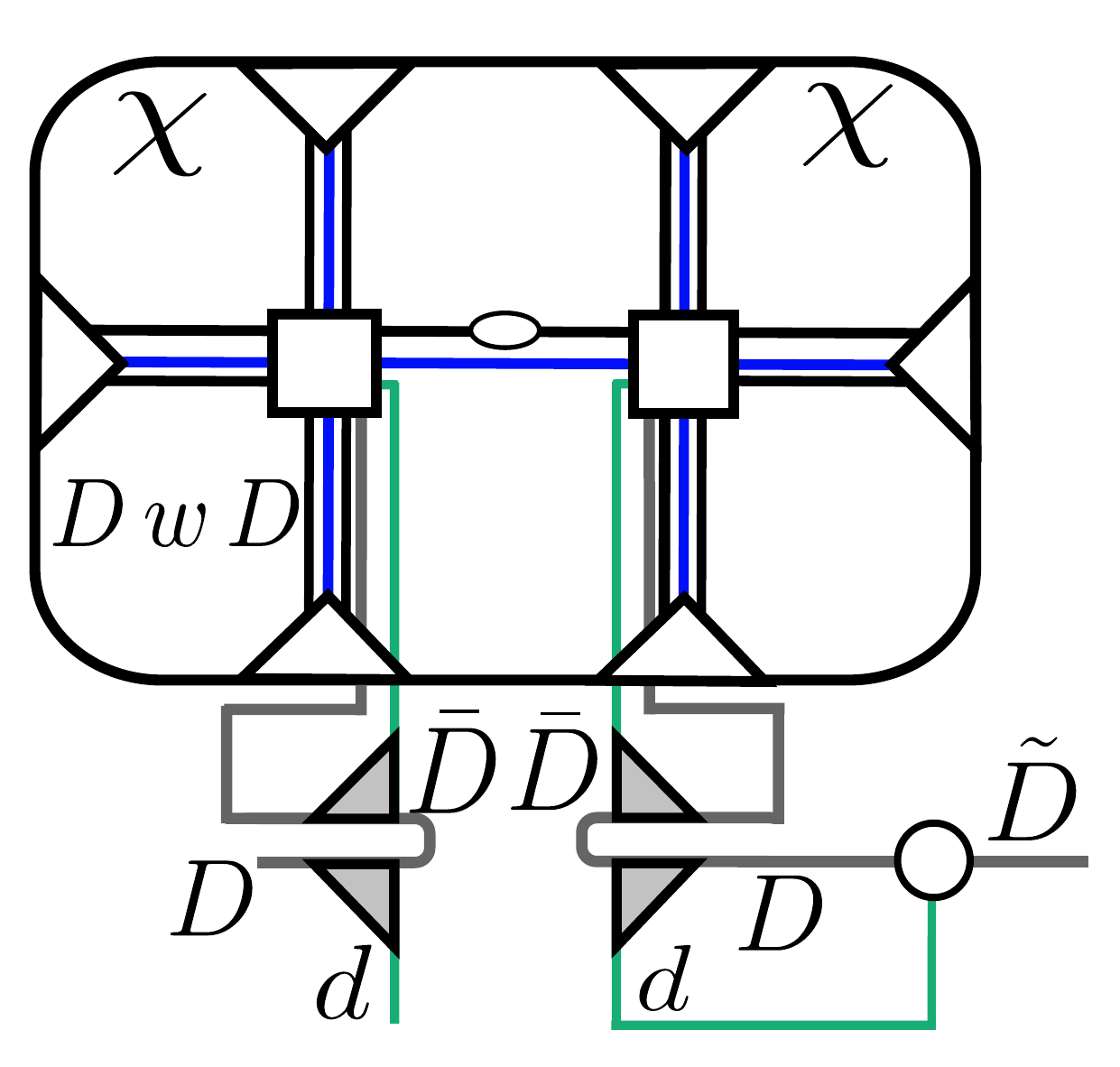}
\centerline{(a)}
\vspace{0.1cm}
\end{minipage}
\hfill
\begin{minipage}[c]{0.2\linewidth}
\includegraphics[width = \textwidth]{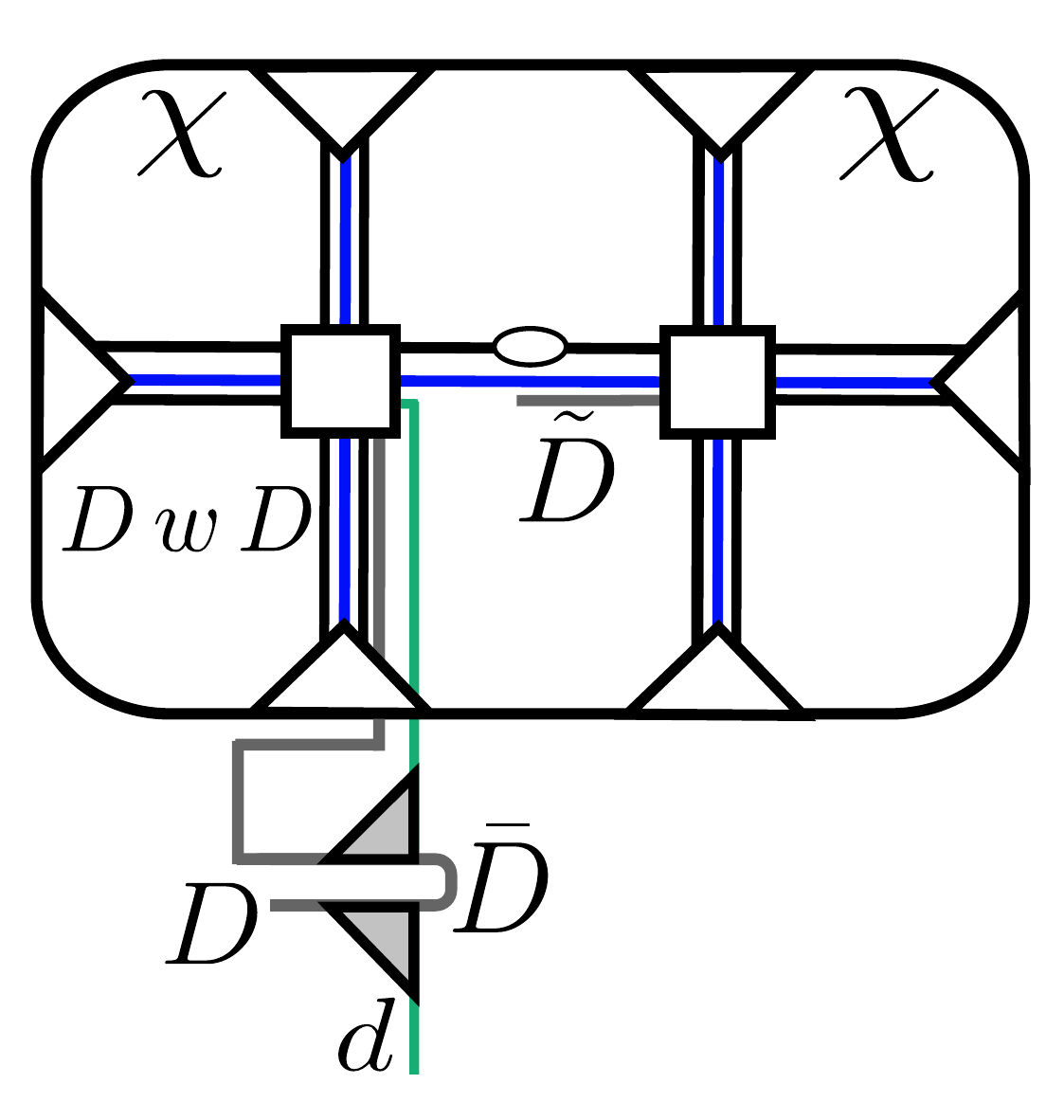}
\centerline{(b)}
\vspace{0.1cm}
\end{minipage}
\hfill
\begin{minipage}[c]{0.15\linewidth}
\includegraphics[width = \textwidth]{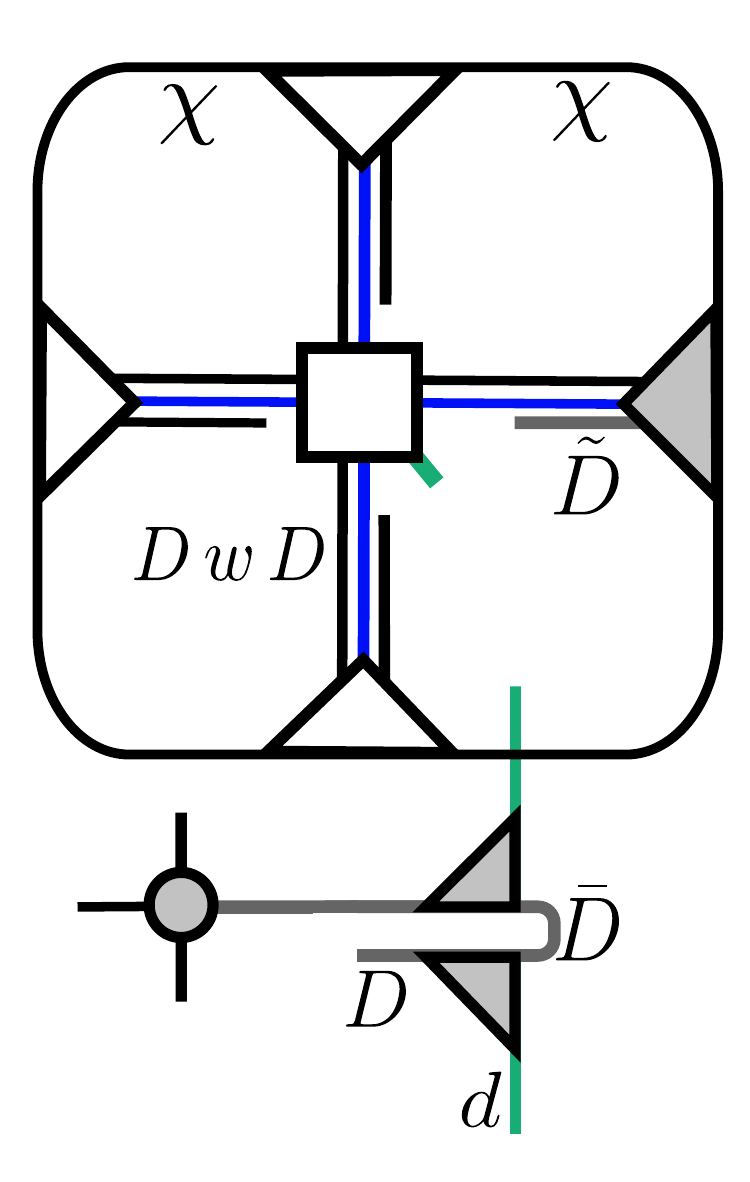}
\centerline{(c)}
\vspace{0.1cm}
\end{minipage}
\hfill
\begin{minipage}[c]{0.15\linewidth}
\includegraphics[width = \textwidth]{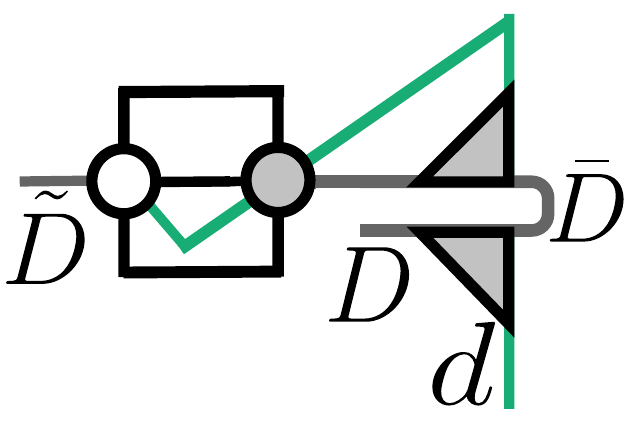}
\centerline{(d)}
\vspace{0.1cm}
\end{minipage}
\hfill
\begin{minipage}[c]{0.08\linewidth}
\includegraphics[width = \textwidth]{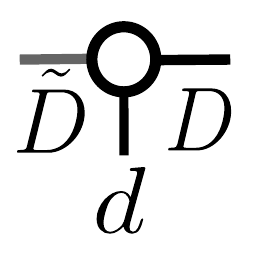}
\centerline{(e)}
\vspace{0.1cm}
\end{minipage}
\label{fig:peps_cbe_steps}
\caption{RSVD for the CBE within energy minimization (a) Initial setup of $A \, \Omega$. (b)
Contraction of right orthogonal projector with $\Omega$. (c) Contraction of four tensors on
the right. (d) Contraction of all tensors modulo the lower PEPS-layer. (e) Contraction of 
remaining tensors.}
\end{figure*}

Fig.~\hyperref[fig:peps_cbe]{6} illustrates the CBE for energy minimization.
The PEPS-tensors are run through the weighted traced gauge \cite{Evenbly2018Aug}, such that
orthogonal projectors can be constructed \cite{Scheb2023Apr}. Otherwise the same arguments apply as
before and we only need to detail the operations of $A \, \Omega$ in 
Fig.~\hyperref[fig:peps_cbe_steps]{7}.
We again start with the initial setup in Fig.~\hyperref[fig:peps_cbe_steps]{7(a)}, where the white circle
in the lower right corner constitutes the Gaussian random matrix $\Omega$.
Since $\chi \gg D,\tilde{D},\bar{D}$, the orthogonal projector for PEPS-tensors can be 
calculated directly and is contracted with $\Omega$ on the right, leading to 
Fig.~\hyperref[fig:peps_cbe_steps]{7(b)}. The right half of the cluster is calculated according to 
Fig.~\hyperref[fig:cont_pattern]{2(a)}, which leaves the tensors illustrated in 
Fig.~\hyperref[fig:peps_cbe_steps]{7(c)}. The PEPS-tensor and its connected projector are dislodged, 
leaving
a cluster that is of the structure $H_{\text{eff}} \left|\psi\right>$ and can therefore be contracted
according to Fig.~\hyperref[fig:cont_pattern]{2(b)}. The remaining, 
computationally inexpensive operations in 
Fig.~\hyperref[fig:peps_cbe_steps]{7(d)} yield the final result in 
Fig.~\hyperref[fig:peps_cbe_steps]{7(e)}.

\section{Results}
\label{sec:results}
Given the computational improvements laid out in the previous chapters, we present benchmark results
for ground state calculations of the two-dimensional Hubbard model. The parameters are 
similar to those used in the latest version of the fPEPS-PEPO algorithm \cite{Scheb2023Apr}: 
Hopping is reduced to nearest neighbours only, onsite repulsion is set to $U=8$ and open boundary 
conditions are implemented. The simulations for the 4$\times$4- and 
6$\times$6-lattices were performed at 
half-filling, 
i.e. $16$ and $36$ electrons, respectively. For the 8$\times$8-lattice,
we chose $1/8$-filling, i.e. 
$56$ electrons, to induce a stripe structure of the local density. As PEPS calculations without
symmetries are intractable in practical simulations, we compared the usage of two different symmetry
groups. For
$\text{U(1)}_{\text{spin}} \otimes \text{U(1)}_{\text{charge}}$ - symmetry, abbreviated as ''U(1)'', 
we picked PEPS bond dimensions ranging from $D=4$ to $D=8$. For
$\text{SU(2)}_{\text{spin}} \otimes \text{U(1)}_{\text{charge}}$ - symmetry, abbreviated as 
''SU(2)'', 
we picked PEPS bond dimensions ranging from $D=4$ to $D=6$. Ground state calculations were performed by
alternating between $3$ local sweeps and $100$ gradient sweeps, 
constituting one supersweep. While the
environment bond dimension ranged from $\chi=250$ to $\chi=400$ in 
Ref.~\cite{Scheb2023Apr},
we were able to increase it to $\chi=500$ for all simulations in this 
paper. The resulting energies
for all lattice sizes were plotted relative to $E_0$, which is the final
energy of a DMRG calculation 
with $D=4000$ states. For the 4$\times$4 lattice, $E_0$ amounts to the 
exact ground state energy, 
whereas for the 6$\times$6 and 8$\times$8 lattice, $E_0$ is an upper 
bound thereof.

\begin{figure}[ht]
\includegraphics[width = 0.45\textwidth]{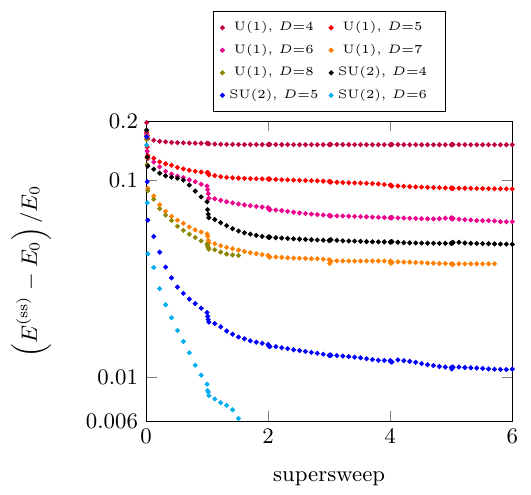}
\label{fig:4x4_energy}
\caption{Relative error in the ground-state energy of the Hubbard
model on a $4 \times 4$ lattice with open boundary conditions, $U=8$,
$S=0$, and $N=16$ (half filling), calculated with fPEPS with U(1)
and SU(2) symmetry.}
\end{figure}

Fig.~\hyperref[fig:4x4_energy]{8} depicts the energy convergence for the 4$\times$4 lattice. 
The curves show a clear variational behaviour, as the energy decreases for higher bond
dimensions and flatten for increasing supersweeps. Several notable differences arise
when comparing them to Fig.~38 in Ref.~\cite{Scheb2023Apr}: First, the (SU(2),$D=4$) energies
now lie between (U(1),$D=6$) and (U(1),$D=7$), whereas previously it seemed to converge to
approximately the same value as (U(1),$D=5$). Second, for both symmetry groups and most 
bond dimensions the energy exhibits a sharp decline at the beginning of the second 
supersweep, which indicates that after the first batch of gradient sweeps a set of local
sweeps has been overdue to optimize the virtual basis between PEPS-tensors. Third,
the (SU(2),$D=6$) penetrates the $1\%$ barrier after one supersweep, whereas it stayed above
it after two supersweeps previously. We attribute all of these differences to the increase 
of $\chi$ from $250$, $300$ and $350$ in Ref.~\cite{Scheb2023Apr} to $500$ in this paper. 
This shows that choosing a $\chi$ that is too low can not just lead to 
numerical instabilities which yield obvious pathological behaviour, but distort the 
convergence in subtle ways that are not detectable in isolation, but only become apparent 
by comparing different values of $\chi$. While the (SU(2),$D=6$) run took 
$4$ days and $150$ GB of memory for $\chi=300$ and $2$ supersweeps, it now takes $5$ days 
and $26$ GB of memory for $\chi=500$ and $1.5$ supersweeps.

\begin{figure}[ht]
\includegraphics[width = 0.45\textwidth]{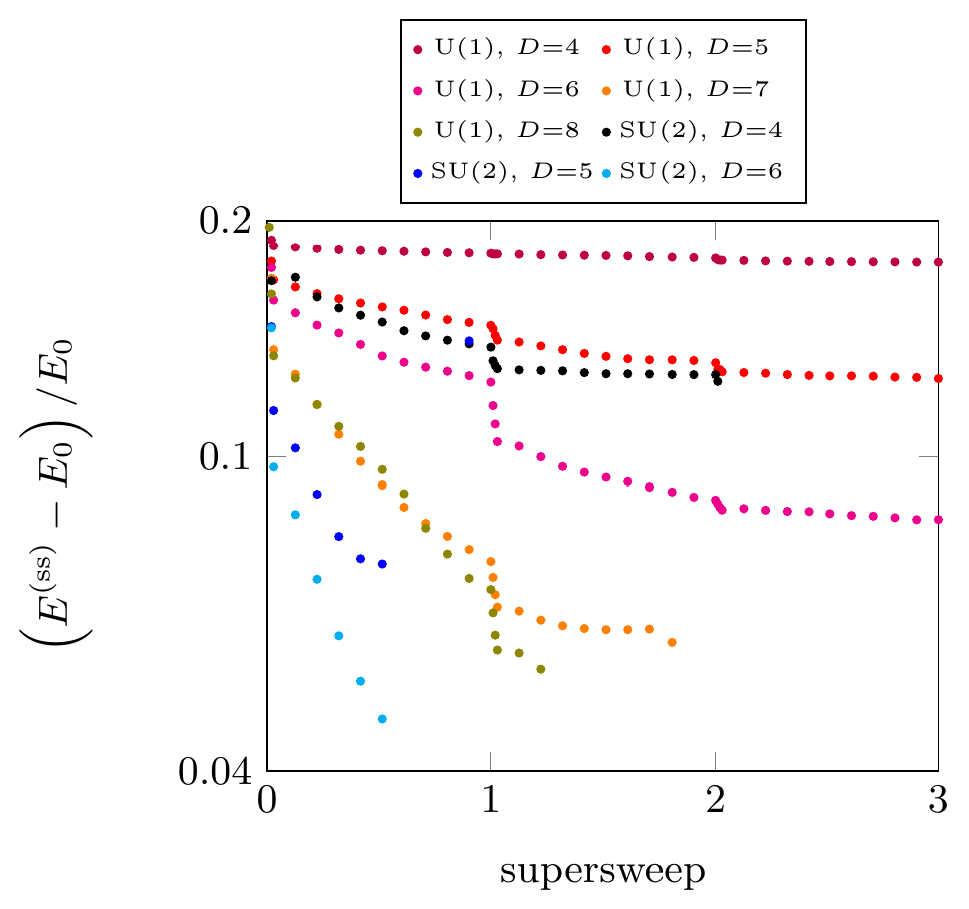}
\label{fig:6x6_energy}
\caption{Relative error in the ground-state energy of the Hubbard
model on a $6 \times 6$ lattice with open boundary conditions, $U=8$,
$S=0$, and $N=36$ (half filling), calculated with fPEPS with U(1)
and SU(2) symmetry.}
\end{figure}

We now proceed to the 6$\times$6 lattice in Fig.~\hyperref[fig:6x6_energy]{9}. 
Again, the energies 
show a clear variational behaviour, although for a lower number of supersweeps due to the
larger system size. Simulations for larger bond dimensions had to be terminated early, as
the algorithm became numerically unstable and rerunning those jobs for higher values of 
$\chi$ was not feasible, even with the computational improvements presented above. The 
drop-off at the beginning of the second supersweep is even more distinct here, indicating
that the proper ratio between the number of local sweeps and the number of gradient sweeps 
ought to be reexamined. The (U(1),$D=7$) curve exhibits a temporary increase at the end, 
pointing to a temporary instability from which the algorithm seems to recover afterwards.
The most striking improvement compared to Fig.~39 in 
Ref.~\cite{Scheb2023Apr} is that we were able to execute a (U(1),$D=8$) simulation, whose
energies approximately coincide with those of (U(1),$D=7$) in the beginning, but then 
become significantly lower in the second supersweep. While the lowest energy error was 
previously $6.8\%$ for (SU(2),$D=6$), we managed to push this number down to $4.7\%$ in 
this paper, albeit at a higher fidelity due to the larger $\chi$ and half the runtime.

\begin{figure}[ht]
\includegraphics[width = 0.45\textwidth]{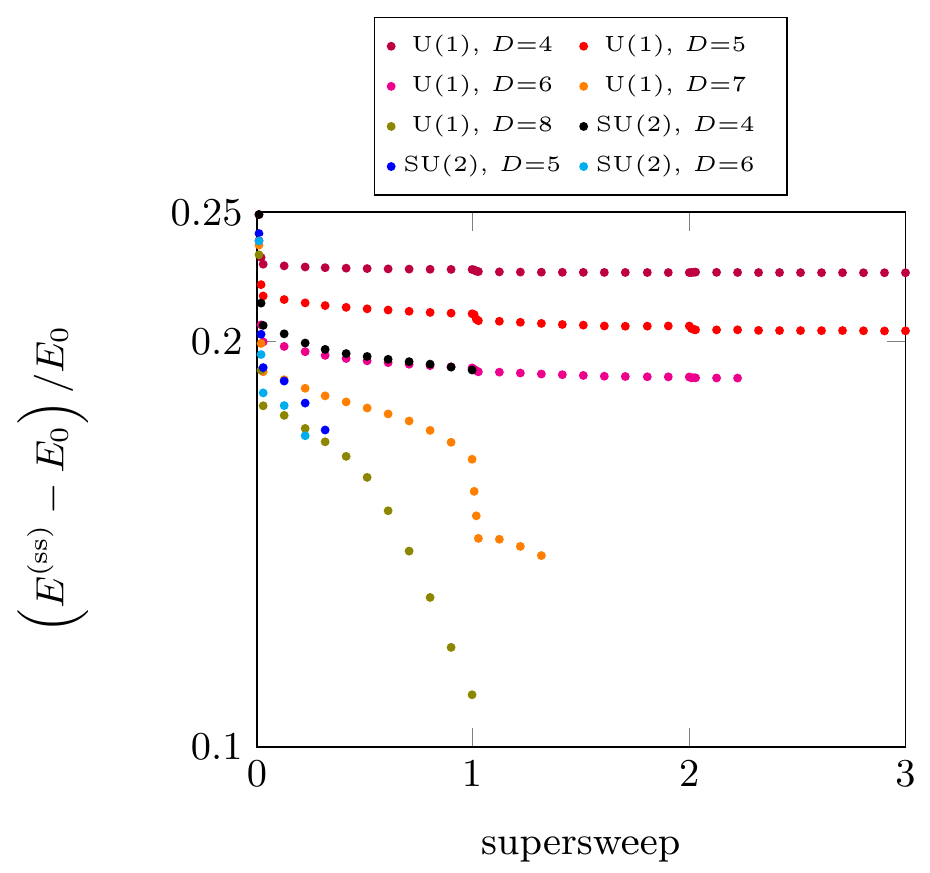}
\label{fig:8x8_energy}
\caption{Relative error in the ground-state energy of the Hubbard
model on a $8 \times 8$ lattice with open boundary conditions, $U=8$,
$S=0$, and $N=56$ (half filling), calculated with fPEPS with U(1)
and SU(2) symmetry.}
\end{figure}

Finally, we comment on the energy convergence of the 8$\times$8 lattice
at $1/8$-filling in 
Fig.~\hyperref[fig:8x8_energy]{10}.
Unfortunately, we were only able to provide a few data points for the SU(2) simulations, as more were
either numerically unstable or took more than two weeks of runtime. Unlike the energies depicted
in Fig.~\hyperref[fig:4x4_energy]{8} and Fig.~\hyperref[fig:6x6_energy]{9}, the lowest energies are 
given by the (U(1),$D=8$) simulation, which after one full supersweep and $8$ days reached a relative
energy error of $11\%$. This stands in contrast to Ref.~\cite{Scheb2023Apr}, where we were only able
to reach an error of $16\%$ after half a supersweep and 21 days.

\begin{figure}[ht]
\includegraphics[width = 0.45\textwidth]{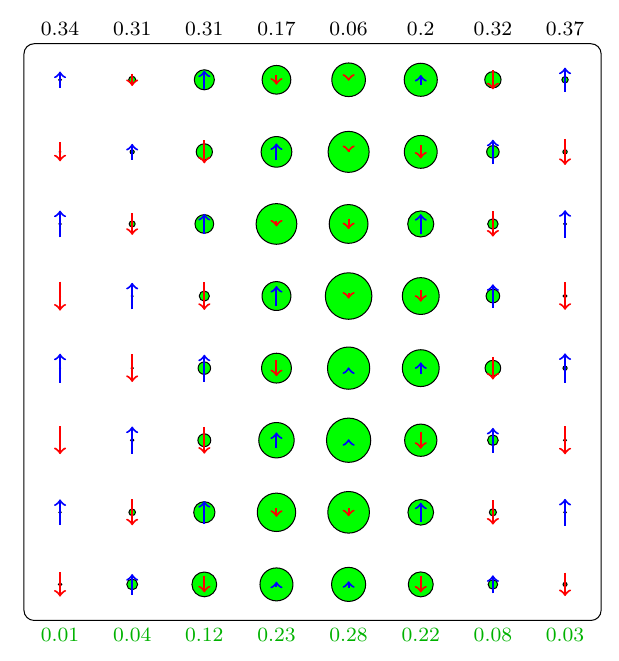}
\label{fig:8x8_U1_den}
\caption{Local $z$-component of the spin $\langle S^z_i \rangle =                       
  \frac{1}{2}\left (n_{i,\uparrow} - n_{i,\downarrow}  \right )$                 
  (size, color, and direction of arrows) and local hole density                  
  $1-\langle n_i \rangle$ (diameter of                                           
  green-shaded circles) on an                                                    
  $8\times 8$ lattice with open boundary conditions calculated with              
  U(1) symmetry, and bond dimension $D=8$.                           
  Here $U=8$, $S_z=0$, and $N=56$ so that $\langle n\rangle = 0.875$.            
  The black numbers are the average $\langle S^z_i \rangle$ for the column of    
  sites below,                                                                   
  and the green numbers on the bottom                                            
  edge are the average hole densities for the column of sites above.}
\end{figure}

To gain some insight into the physical behaviour of the Hubbard model at 1/8 filling, we also present
the local density for the U(1) symmetric case at bond dimension $8$ in 
Fig.~\hyperref[fig:8x8_U1_den]{11}.
The $z$-component of the spin is depicted as a blue arrow for positive values and a red arrow for
negative values. The hole density are represented by green circles. 
As expected, we observe the well-known stripe
structure \cite{Zheng2017Dec} of an oscillating charge density, combined with incommensurate 
antiferromagnetism. The charge oscillation is edge-centered at the top and bottom of the lattice,
but appears to be site-centered in the middle. Since the algorithm is far from converged, we are
unable to determine whether this behaviour is closer to the actual physical setup of an 8$\times$8 lattice
with open boundary conditions, or a numerical artefact.

We also present the local density for the SU(2) symmetric case in Fig.~\hyperref[fig:8x8_U1_den]{12}.
\begin{figure}[ht]
\includegraphics[width = 0.45\textwidth]{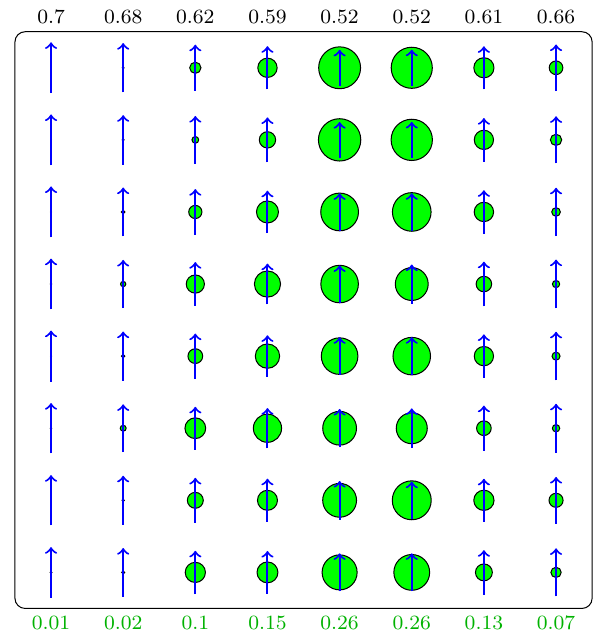}
\label{fig:8x8_SU2_den}
\caption{
  Local spin density $\langle\mathbf{S}^2_i\rangle$ (size of                     
  blue arrows)                                                                   
  and local hole density $1 - \langle n_i \rangle$ (diameter of                  
  green-shaded circles)  for the Hubbard model on an $8\times 8$                 
  lattice with open boundary conditions and                                      
  $U=8$, $S=0$, and $N=56$ so that $\langle n\rangle = 0.875$,                   
  calculated with SU(2) symmetry, and bond dimension $D=6$.          
  The black numbers are the average spin density for the column of               
  sites below,                                                                       
  and the green numbers on the bottom                                            
  edge are the average hole densities for the column of sites above.}
\end{figure}
Since the $z$-component of the spin is zero by construction and antiferromagnetic order is suppressed, 
we depict the total spin component instead. The charge density exhibits a similar stripe structure
as in the U(1) case, whose distribution is a qualitative improvement over the less symmetric 
distribution in Ref.~\cite{Scheb2023Apr}. We again note that the simulation is far from converged 
and one should therefore expect the hole density to shift significantly for a more progressed 
simulation.

\section{Summary and Outlook}
\label{sec:SummaryOutlook}
In this paper, we explained in detail how to contract finite PEPSs without any gauge 
constraints. The first technical section (Sec.~\ref{sec:optcont}) concerned itself with the optimal 
contraction of the two
dominant contraction patterns and how to slice them such that
memory usage remains minimal. The second technical chapter (Sec.~\ref{sec:cbersvd}) illustrated how to 
combine the CBE with the RSVD and apply this factorization framework to the fPEPS algorithm. 
Finally, we provided some benchmark results in Sec.~\ref{sec:results} and compared them to the 
previous version of the fPEPS framework in Ref.~\cite{Scheb2023Apr}. For all three system sizes, 
we were able to reach lower energies, at a higher $\chi$ and lower runtime, therefor justifying the
technical improvements presented.

However, as is evident from the data, even these improvements did not yield energies that come close
to the upper bounds provided by the DMRG, meaning that in its current form, the fPEPS-PEPO scheme
is still not a competitive tool for calculating two-dimensional quantum systems. For future research, 
the ideas presented in this paper and its precursor \cite{Scheb2023Apr} have to be combined with other
lines of inquiry. One promising option is to incorporate them into the contraction of fPEPS via 
Monte Carlo methods \cite{Vieijra2021Dec,Lin2024Jun,Liu2025Jun}, which allow for a more efficient 
tensor contraction at the 
price of another error. Another possibility would be to apply the recently developed 
belief propagation (BP) \cite{Tindall2023Dec}, which makes an attempt at canonicalizing
cyclic tensor networks. Since BP is not guaranteed to converge to an optimal 
wavefunction for a given bond dimension, 
neglected contributions can be reincorporated through a loop series 
expansion \cite{Evenbly2024Sep}. It remains an open question whether these or 
other potentially new 
improvements will yield a breakthrough and whether fPEPSs have the practical capacity
to describe arbitrary large, heterogeneous, two-dimensional quantum systems.

\begin{acknowledgments}
We thank Jan von Delft, Andreas Gleis, Jheng-Wei Li and Yuan Gao for fruitful discussions. 
This work was funded in part by the Deutsche Forschungsgemeinschaft under Germany’s Excellence Strategy EXC-
2111 (Project No. 390814868). It is part of the Munich
Quantum Valley, supported by the Bavarian state government 
with funds from the Hightech Agenda Bayern Plus.
\end{acknowledgments}

\FloatBarrier

\bibliography{paper_bib}

\begin{thebibliography}{42}%
\makeatletter
\providecommand \@ifxundefined [1]{%
 \@ifx{#1\undefined}
}%
\providecommand \@ifnum [1]{%
 \ifnum #1\expandafter \@firstoftwo
 \else \expandafter \@secondoftwo
 \fi
}%
\providecommand \@ifx [1]{%
 \ifx #1\expandafter \@firstoftwo
 \else \expandafter \@secondoftwo
 \fi
}%
\providecommand \natexlab [1]{#1}%
\providecommand \enquote  [1]{``#1''}%
\providecommand \bibnamefont  [1]{#1}%
\providecommand \bibfnamefont [1]{#1}%
\providecommand \citenamefont [1]{#1}%
\providecommand \href@noop [0]{\@secondoftwo}%
\providecommand \href [0]{\begingroup \@sanitize@url \@href}%
\providecommand \@href[1]{\@@startlink{#1}\@@href}%
\providecommand \@@href[1]{\endgroup#1\@@endlink}%
\providecommand \@sanitize@url [0]{\catcode `\\12\catcode `\$12\catcode
  `\&12\catcode `\#12\catcode `\^12\catcode `\_12\catcode `\%12\relax}%
\providecommand \@@startlink[1]{}%
\providecommand \@@endlink[0]{}%
\providecommand \url  [0]{\begingroup\@sanitize@url \@url }%
\providecommand \@url [1]{\endgroup\@href {#1}{\urlprefix }}%
\providecommand \urlprefix  [0]{URL }%
\providecommand \Eprint [0]{\href }%
\providecommand \doibase [0]{https://doi.org/}%
\providecommand \selectlanguage [0]{\@gobble}%
\providecommand \bibinfo  [0]{\@secondoftwo}%
\providecommand \bibfield  [0]{\@secondoftwo}%
\providecommand \translation [1]{[#1]}%
\providecommand \BibitemOpen [0]{}%
\providecommand \bibitemStop [0]{}%
\providecommand \bibitemNoStop [0]{.\EOS\space}%
\providecommand \EOS [0]{\spacefactor3000\relax}%
\providecommand \BibitemShut  [1]{\csname bibitem#1\endcsname}%
\let\auto@bib@innerbib\@empty
\bibitem [{\citenamefont
  {Or{\ifmmode\acute{u}\else\'{u}\fi}s}(2019)}]{Orus2019Sep}%
  \BibitemOpen
  \bibfield  {author} {\bibinfo {author} {\bibfnamefont {R.}~\bibnamefont
  {Or{\ifmmode\acute{u}\else\'{u}\fi}s}},\ }\bibfield  {title} {\bibinfo
  {title} {{Tensor networks for complex quantum systems}},\ }\href
  {https://doi.org/10.1038/s42254-019-0086-7} {\bibfield  {journal} {\bibinfo
  {journal} {Nat. Rev. Phys.}\ }\textbf {\bibinfo {volume} {1}},\ \bibinfo
  {pages} {538} (\bibinfo {year} {2019})}\BibitemShut {NoStop}%
\bibitem [{\citenamefont {Cirac}\ \emph {et~al.}(2021)\citenamefont {Cirac},
  \citenamefont
  {P{\ifmmode\acute{e}\else\'{e}\fi}rez-Garc{\ifmmode\acute{\imath}\else\'{\i}\fi}a},
  \citenamefont {Schuch},\ and\ \citenamefont {Verstraete}}]{Cirac2021Dec}%
  \BibitemOpen
  \bibfield  {author} {\bibinfo {author} {\bibfnamefont {J.~I.}\ \bibnamefont
  {Cirac}}, \bibinfo {author} {\bibfnamefont {D.}~\bibnamefont
  {P{\ifmmode\acute{e}\else\'{e}\fi}rez-Garc{\ifmmode\acute{\imath}\else\'{\i}\fi}a}},
  \bibinfo {author} {\bibfnamefont {N.}~\bibnamefont {Schuch}},\ and\ \bibinfo
  {author} {\bibfnamefont {F.}~\bibnamefont {Verstraete}},\ }\bibfield  {title}
  {\bibinfo {title} {{Matrix product states and projected entangled pair
  states: Concepts, symmetries, theorems}},\ }\href
  {https://doi.org/10.1103/RevModPhys.93.045003} {\bibfield  {journal}
  {\bibinfo  {journal} {Rev. Mod. Phys.}\ }\textbf {\bibinfo {volume} {93}},\
  \bibinfo {pages} {045003} (\bibinfo {year} {2021})}\BibitemShut {NoStop}%
\bibitem [{\citenamefont
  {Ba{\ifmmode\tilde{n}\else\~{n}\fi}uls}(2023)}]{Banuls2023Mar}%
  \BibitemOpen
  \bibfield  {author} {\bibinfo {author} {\bibfnamefont {M.~C.}\ \bibnamefont
  {Ba{\ifmmode\tilde{n}\else\~{n}\fi}uls}},\ }\bibfield  {title} {\bibinfo
  {title} {{Tensor Network Algorithms: A Route Map}},\ }\href
  {https://doi.org/10.1146/annurev-conmatphys-040721-022705} {\bibfield
  {journal} {\bibinfo  {journal} {Annu. Rev. Condens. Matter Phys.}\ ,\
  \bibinfo {pages} {173}} (\bibinfo {year} {2023})}\BibitemShut {NoStop}%
\bibitem [{\citenamefont {White}(1992)}]{White1992Nov}%
  \BibitemOpen
  \bibfield  {author} {\bibinfo {author} {\bibfnamefont {S.~R.}\ \bibnamefont
  {White}},\ }\bibfield  {title} {\bibinfo {title} {{Density matrix formulation
  for quantum renormalization groups}},\ }\href
  {https://doi.org/10.1103/PhysRevLett.69.2863} {\bibfield  {journal} {\bibinfo
   {journal} {Phys. Rev. Lett.}\ }\textbf {\bibinfo {volume} {69}},\ \bibinfo
  {pages} {2863} (\bibinfo {year} {1992})}\BibitemShut {NoStop}%
\bibitem [{\citenamefont {White}(1993)}]{White1993Oct}%
  \BibitemOpen
  \bibfield  {author} {\bibinfo {author} {\bibfnamefont {S.~R.}\ \bibnamefont
  {White}},\ }\bibfield  {title} {\bibinfo {title} {{Density-matrix algorithms
  for quantum renormalization groups}},\ }\href
  {https://doi.org/10.1103/PhysRevB.48.10345} {\bibfield  {journal} {\bibinfo
  {journal} {Phys. Rev. B}\ }\textbf {\bibinfo {volume} {48}},\ \bibinfo
  {pages} {10345} (\bibinfo {year} {1993})}\BibitemShut {NoStop}%
\bibitem [{\citenamefont {Rommer}\ and\ \citenamefont
  {{\ifmmode\ddot{O}\else\"{O}\fi}stlund}(1997)}]{Rommer1997Jan}%
  \BibitemOpen
  \bibfield  {author} {\bibinfo {author} {\bibfnamefont {S.}~\bibnamefont
  {Rommer}}\ and\ \bibinfo {author} {\bibfnamefont {S.}~\bibnamefont
  {{\ifmmode\ddot{O}\else\"{O}\fi}stlund}},\ }\bibfield  {title} {\bibinfo
  {title} {{Class of ansatz wave functions for one-dimensional spin systems and
  their relation to the density matrix renormalization group}},\ }\href
  {https://doi.org/10.1103/PhysRevB.55.2164} {\bibfield  {journal} {\bibinfo
  {journal} {Phys. Rev. B}\ }\textbf {\bibinfo {volume} {55}},\ \bibinfo
  {pages} {2164} (\bibinfo {year} {1997})}\BibitemShut {NoStop}%
\bibitem [{\citenamefont
  {Schollw{\ifmmode\ddot{o}\else\"{o}\fi}ck}(2011)}]{Schollwock2011Jan}%
  \BibitemOpen
  \bibfield  {author} {\bibinfo {author} {\bibfnamefont {U.}~\bibnamefont
  {Schollw{\ifmmode\ddot{o}\else\"{o}\fi}ck}},\ }\bibfield  {title} {\bibinfo
  {title} {{The density-matrix renormalization group in the age of matrix
  product states}},\ }\href {https://doi.org/10.1016/j.aop.2010.09.012}
  {\bibfield  {journal} {\bibinfo  {journal} {Ann. Phys.}\ }\textbf {\bibinfo
  {volume} {326}},\ \bibinfo {pages} {96} (\bibinfo {year} {2011})}\BibitemShut
  {NoStop}%
\bibitem [{\citenamefont {Verstraete}\ and\ \citenamefont
  {Cirac}(2004)}]{Verstraete2004Jul}%
  \BibitemOpen
  \bibfield  {author} {\bibinfo {author} {\bibfnamefont {F.}~\bibnamefont
  {Verstraete}}\ and\ \bibinfo {author} {\bibfnamefont {J.~I.}\ \bibnamefont
  {Cirac}},\ }\bibfield  {title} {\bibinfo {title} {{Renormalization algorithms
  for Quantum-Many Body Systems in two and higher dimensions}},\ }\bibfield
  {journal} {\bibinfo  {journal} {arXiv}\ }\href
  {https://doi.org/10.48550/arXiv.cond-mat/0407066}
  {10.48550/arXiv.cond-mat/0407066} (\bibinfo {year} {2004}),\ \Eprint
  {https://arxiv.org/abs/cond-mat/0407066} {cond-mat/0407066} \BibitemShut
  {NoStop}%
\bibitem [{\citenamefont {Perez-Garcia}\ \emph {et~al.}(2007)\citenamefont
  {Perez-Garcia}, \citenamefont {Verstraete}, \citenamefont {Cirac},\ and\
  \citenamefont {Wolf}}]{Perez-Garcia2007Jul}%
  \BibitemOpen
  \bibfield  {author} {\bibinfo {author} {\bibfnamefont {D.}~\bibnamefont
  {Perez-Garcia}}, \bibinfo {author} {\bibfnamefont {F.}~\bibnamefont
  {Verstraete}}, \bibinfo {author} {\bibfnamefont {J.~I.}\ \bibnamefont
  {Cirac}},\ and\ \bibinfo {author} {\bibfnamefont {M.~M.}\ \bibnamefont
  {Wolf}},\ }\bibfield  {title} {\bibinfo {title} {{PEPS as unique ground
  states of local Hamiltonians}},\ }\bibfield  {journal} {\bibinfo  {journal}
  {arXiv}\ }\href {https://doi.org/10.48550/arXiv.0707.2260}
  {10.48550/arXiv.0707.2260} (\bibinfo {year} {2007}),\ \Eprint
  {https://arxiv.org/abs/0707.2260} {0707.2260} \BibitemShut {NoStop}%
\bibitem [{\citenamefont {Schuch}\ \emph {et~al.}(2010)\citenamefont {Schuch},
  \citenamefont {Cirac},\ and\ \citenamefont
  {P{\ifmmode\acute{e}\else\'{e}\fi}rez-Garc{\ifmmode\acute{\imath}\else\'{\i}\fi}a}}]{Schuch2010Oct}%
  \BibitemOpen
  \bibfield  {author} {\bibinfo {author} {\bibfnamefont {N.}~\bibnamefont
  {Schuch}}, \bibinfo {author} {\bibfnamefont {I.}~\bibnamefont {Cirac}},\ and\
  \bibinfo {author} {\bibfnamefont {D.}~\bibnamefont
  {P{\ifmmode\acute{e}\else\'{e}\fi}rez-Garc{\ifmmode\acute{\imath}\else\'{\i}\fi}a}},\
  }\bibfield  {title} {\bibinfo {title} {{PEPS as ground states: Degeneracy and
  topology}},\ }\href {https://doi.org/10.1016/j.aop.2010.05.008} {\bibfield
  {journal} {\bibinfo  {journal} {Ann. Phys.}\ }\textbf {\bibinfo {volume}
  {325}},\ \bibinfo {pages} {2153} (\bibinfo {year} {2010})}\BibitemShut
  {NoStop}%
\bibitem [{\citenamefont {Verstraete}\ \emph {et~al.}(2006)\citenamefont
  {Verstraete}, \citenamefont {Wolf}, \citenamefont {Perez-Garcia},\ and\
  \citenamefont {Cirac}}]{Verstraete2006Jun}%
  \BibitemOpen
  \bibfield  {author} {\bibinfo {author} {\bibfnamefont {F.}~\bibnamefont
  {Verstraete}}, \bibinfo {author} {\bibfnamefont {M.~M.}\ \bibnamefont
  {Wolf}}, \bibinfo {author} {\bibfnamefont {D.}~\bibnamefont {Perez-Garcia}},\
  and\ \bibinfo {author} {\bibfnamefont {J.~I.}\ \bibnamefont {Cirac}},\
  }\bibfield  {title} {\bibinfo {title} {{Criticality, the Area Law, and the
  Computational Power of Projected Entangled Pair States}},\ }\href
  {https://doi.org/10.1103/PhysRevLett.96.220601} {\bibfield  {journal}
  {\bibinfo  {journal} {Phys. Rev. Lett.}\ }\textbf {\bibinfo {volume} {96}},\
  \bibinfo {pages} {220601} (\bibinfo {year} {2006})}\BibitemShut {NoStop}%
\bibitem [{\citenamefont {Jordan}\ \emph {et~al.}(2008)\citenamefont {Jordan},
  \citenamefont {Or{\ifmmode\acute{u}\else\'{u}\fi}s}, \citenamefont {Vidal},
  \citenamefont {Verstraete},\ and\ \citenamefont {Cirac}}]{Jordan2008Dec}%
  \BibitemOpen
  \bibfield  {author} {\bibinfo {author} {\bibfnamefont {J.}~\bibnamefont
  {Jordan}}, \bibinfo {author} {\bibfnamefont {R.}~\bibnamefont
  {Or{\ifmmode\acute{u}\else\'{u}\fi}s}}, \bibinfo {author} {\bibfnamefont
  {G.}~\bibnamefont {Vidal}}, \bibinfo {author} {\bibfnamefont
  {F.}~\bibnamefont {Verstraete}},\ and\ \bibinfo {author} {\bibfnamefont
  {J.~I.}\ \bibnamefont {Cirac}},\ }\bibfield  {title} {\bibinfo {title}
  {{Classical Simulation of Infinite-Size Quantum Lattice Systems in Two
  Spatial Dimensions}},\ }\href
  {https://doi.org/10.1103/PhysRevLett.101.250602} {\bibfield  {journal}
  {\bibinfo  {journal} {Phys. Rev. Lett.}\ }\textbf {\bibinfo {volume} {101}},\
  \bibinfo {pages} {250602} (\bibinfo {year} {2008})}\BibitemShut {NoStop}%
\bibitem [{\citenamefont {Corboz}(2016{\natexlab{a}})}]{Corboz2016Jul}%
  \BibitemOpen
  \bibfield  {author} {\bibinfo {author} {\bibfnamefont {P.}~\bibnamefont
  {Corboz}},\ }\bibfield  {title} {\bibinfo {title} {{Variational optimization
  with infinite projected entangled-pair states}},\ }\href
  {https://doi.org/10.1103/PhysRevB.94.035133} {\bibfield  {journal} {\bibinfo
  {journal} {Phys. Rev. B}\ }\textbf {\bibinfo {volume} {94}},\ \bibinfo
  {pages} {035133} (\bibinfo {year} {2016}{\natexlab{a}})}\BibitemShut
  {NoStop}%
\bibitem [{\citenamefont {Phien}\ \emph {et~al.}(2015)\citenamefont {Phien},
  \citenamefont {Bengua}, \citenamefont {Tuan}, \citenamefont {Corboz},\ and\
  \citenamefont {Or{\ifmmode\acute{u}\else\'{u}\fi}s}}]{Phien2015Jul}%
  \BibitemOpen
  \bibfield  {author} {\bibinfo {author} {\bibfnamefont {H.~N.}\ \bibnamefont
  {Phien}}, \bibinfo {author} {\bibfnamefont {J.~A.}\ \bibnamefont {Bengua}},
  \bibinfo {author} {\bibfnamefont {H.~D.}\ \bibnamefont {Tuan}}, \bibinfo
  {author} {\bibfnamefont {P.}~\bibnamefont {Corboz}},\ and\ \bibinfo {author}
  {\bibfnamefont {R.}~\bibnamefont {Or{\ifmmode\acute{u}\else\'{u}\fi}s}},\
  }\bibfield  {title} {\bibinfo {title} {{Infinite projected entangled pair
  states algorithm improved: Fast full update and gauge fixing}},\ }\href
  {https://doi.org/10.1103/PhysRevB.92.035142} {\bibfield  {journal} {\bibinfo
  {journal} {Phys. Rev. B}\ }\textbf {\bibinfo {volume} {92}},\ \bibinfo
  {pages} {035142} (\bibinfo {year} {2015})}\BibitemShut {NoStop}%
\bibitem [{\citenamefont {Corboz}\ \emph {et~al.}(2011)\citenamefont {Corboz},
  \citenamefont {White}, \citenamefont {Vidal},\ and\ \citenamefont
  {Troyer}}]{Corboz2011Jul}%
  \BibitemOpen
  \bibfield  {author} {\bibinfo {author} {\bibfnamefont {P.}~\bibnamefont
  {Corboz}}, \bibinfo {author} {\bibfnamefont {S.~R.}\ \bibnamefont {White}},
  \bibinfo {author} {\bibfnamefont {G.}~\bibnamefont {Vidal}},\ and\ \bibinfo
  {author} {\bibfnamefont {M.}~\bibnamefont {Troyer}},\ }\bibfield  {title}
  {\bibinfo {title} {{Stripes in the two-dimensional $t$-$J$ model with
  infinite projected entangled-pair states}},\ }\href
  {https://doi.org/10.1103/PhysRevB.84.041108} {\bibfield  {journal} {\bibinfo
  {journal} {Phys. Rev. B}\ }\textbf {\bibinfo {volume} {84}},\ \bibinfo
  {pages} {041108} (\bibinfo {year} {2011})}\BibitemShut {NoStop}%
\bibitem [{\citenamefont {Corboz}(2016{\natexlab{b}})}]{Corboz2016Jan}%
  \BibitemOpen
  \bibfield  {author} {\bibinfo {author} {\bibfnamefont {P.}~\bibnamefont
  {Corboz}},\ }\bibfield  {title} {\bibinfo {title} {{Improved energy
  extrapolation with infinite projected entangled-pair states applied to the
  two-dimensional Hubbard model}},\ }\href
  {https://doi.org/10.1103/PhysRevB.93.045116} {\bibfield  {journal} {\bibinfo
  {journal} {Phys. Rev. B}\ }\textbf {\bibinfo {volume} {93}},\ \bibinfo
  {pages} {045116} (\bibinfo {year} {2016}{\natexlab{b}})}\BibitemShut
  {NoStop}%
\bibitem [{\citenamefont {Zhang}\ \emph {et~al.}(2025)\citenamefont {Zhang},
  \citenamefont {Li}, \citenamefont {Nikolaidou},\ and\ \citenamefont {von
  Delft}}]{Zhang2025Mar}%
  \BibitemOpen
  \bibfield  {author} {\bibinfo {author} {\bibfnamefont {C.}~\bibnamefont
  {Zhang}}, \bibinfo {author} {\bibfnamefont {J.-W.}\ \bibnamefont {Li}},
  \bibinfo {author} {\bibfnamefont {D.}~\bibnamefont {Nikolaidou}},\ and\
  \bibinfo {author} {\bibfnamefont {J.}~\bibnamefont {von Delft}},\ }\bibfield
  {title} {\bibinfo {title} {{Frustration-Induced Superconductivity in the
  $t\text{\ensuremath{-}}{t}^{\ensuremath{'}}$ Hubbard Model}},\ }\href
  {https://doi.org/10.1103/PhysRevLett.134.116502} {\bibfield  {journal}
  {\bibinfo  {journal} {Phys. Rev. Lett.}\ }\textbf {\bibinfo {volume} {134}},\
  \bibinfo {pages} {116502} (\bibinfo {year} {2025})}\BibitemShut {NoStop}%
\bibitem [{\citenamefont {Li}\ \emph {et~al.}(2021)\citenamefont {Li},
  \citenamefont {Bruognolo}, \citenamefont {Weichselbaum},\ and\ \citenamefont
  {von Delft}}]{Li2021Feb}%
  \BibitemOpen
  \bibfield  {author} {\bibinfo {author} {\bibfnamefont {J.-W.}\ \bibnamefont
  {Li}}, \bibinfo {author} {\bibfnamefont {B.}~\bibnamefont {Bruognolo}},
  \bibinfo {author} {\bibfnamefont {A.}~\bibnamefont {Weichselbaum}},\ and\
  \bibinfo {author} {\bibfnamefont {J.}~\bibnamefont {von Delft}},\ }\bibfield
  {title} {\bibinfo {title} {{Study of spin symmetry in the doped
  $t\ensuremath{-}J$ model using infinite projected entangled pair states}},\
  }\href {https://doi.org/10.1103/PhysRevB.103.075127} {\bibfield  {journal}
  {\bibinfo  {journal} {Phys. Rev. B}\ }\textbf {\bibinfo {volume} {103}},\
  \bibinfo {pages} {075127} (\bibinfo {year} {2021})}\BibitemShut {NoStop}%
\bibitem [{\citenamefont {Zaletel}\ and\ \citenamefont
  {Pollmann}(2020)}]{Zaletel2020Jan}%
  \BibitemOpen
  \bibfield  {author} {\bibinfo {author} {\bibfnamefont {M.~P.}\ \bibnamefont
  {Zaletel}}\ and\ \bibinfo {author} {\bibfnamefont {F.}~\bibnamefont
  {Pollmann}},\ }\bibfield  {title} {\bibinfo {title} {{Isometric Tensor
  Network States in Two Dimensions}},\ }\href
  {https://doi.org/10.1103/PhysRevLett.124.037201} {\bibfield  {journal}
  {\bibinfo  {journal} {Phys. Rev. Lett.}\ }\textbf {\bibinfo {volume} {124}},\
  \bibinfo {pages} {037201} (\bibinfo {year} {2020})}\BibitemShut {NoStop}%
\bibitem [{\citenamefont {Soejima}\ \emph {et~al.}(2020)\citenamefont
  {Soejima}, \citenamefont {Siva}, \citenamefont {Bultinck}, \citenamefont
  {Chatterjee}, \citenamefont {Pollmann},\ and\ \citenamefont
  {Zaletel}}]{Soejima2020Feb}%
  \BibitemOpen
  \bibfield  {author} {\bibinfo {author} {\bibfnamefont {T.}~\bibnamefont
  {Soejima}}, \bibinfo {author} {\bibfnamefont {K.}~\bibnamefont {Siva}},
  \bibinfo {author} {\bibfnamefont {N.}~\bibnamefont {Bultinck}}, \bibinfo
  {author} {\bibfnamefont {S.}~\bibnamefont {Chatterjee}}, \bibinfo {author}
  {\bibfnamefont {F.}~\bibnamefont {Pollmann}},\ and\ \bibinfo {author}
  {\bibfnamefont {M.~P.}\ \bibnamefont {Zaletel}},\ }\bibfield  {title}
  {\bibinfo {title} {{Isometric tensor network representation of string-net
  liquids}},\ }\href {https://doi.org/10.1103/PhysRevB.101.085117} {\bibfield
  {journal} {\bibinfo  {journal} {Phys. Rev. B}\ }\textbf {\bibinfo {volume}
  {101}},\ \bibinfo {pages} {085117} (\bibinfo {year} {2020})}\BibitemShut
  {NoStop}%
\bibitem [{\citenamefont {Lin}\ \emph {et~al.}(2022)\citenamefont {Lin},
  \citenamefont {Zaletel},\ and\ \citenamefont {Pollmann}}]{Lin2022Dec}%
  \BibitemOpen
  \bibfield  {author} {\bibinfo {author} {\bibfnamefont {S.-H.}\ \bibnamefont
  {Lin}}, \bibinfo {author} {\bibfnamefont {M.~P.}\ \bibnamefont {Zaletel}},\
  and\ \bibinfo {author} {\bibfnamefont {F.}~\bibnamefont {Pollmann}},\
  }\bibfield  {title} {\bibinfo {title} {{Efficient simulation of dynamics in
  two-dimensional quantum spin systems with isometric tensor networks}},\
  }\href {https://doi.org/10.1103/PhysRevB.106.245102} {\bibfield  {journal}
  {\bibinfo  {journal} {Phys. Rev. B}\ }\textbf {\bibinfo {volume} {106}},\
  \bibinfo {pages} {245102} (\bibinfo {year} {2022})}\BibitemShut {NoStop}%
\bibitem [{\citenamefont {Kadow}\ \emph {et~al.}(2023)\citenamefont {Kadow},
  \citenamefont {Pollmann},\ and\ \citenamefont {Knap}}]{Kadow2023May}%
  \BibitemOpen
  \bibfield  {author} {\bibinfo {author} {\bibfnamefont {W.}~\bibnamefont
  {Kadow}}, \bibinfo {author} {\bibfnamefont {F.}~\bibnamefont {Pollmann}},\
  and\ \bibinfo {author} {\bibfnamefont {M.}~\bibnamefont {Knap}},\ }\bibfield
  {title} {\bibinfo {title} {{Isometric tensor network representations of
  two-dimensional thermal states}},\ }\href
  {https://doi.org/10.1103/PhysRevB.107.205106} {\bibfield  {journal} {\bibinfo
   {journal} {Phys. Rev. B}\ }\textbf {\bibinfo {volume} {107}},\ \bibinfo
  {pages} {205106} (\bibinfo {year} {2023})}\BibitemShut {NoStop}%
\bibitem [{\citenamefont {Emonts}\ and\ \citenamefont
  {Zohar}(2023)}]{Emonts2023Jul}%
  \BibitemOpen
  \bibfield  {author} {\bibinfo {author} {\bibfnamefont {P.}~\bibnamefont
  {Emonts}}\ and\ \bibinfo {author} {\bibfnamefont {E.}~\bibnamefont {Zohar}},\
  }\bibfield  {title} {\bibinfo {title} {{Fermionic Gaussian projected
  entangled pair states in $3+1\mathrm{D}$: Rotations and relativistic
  limits}},\ }\href {https://doi.org/10.1103/PhysRevD.108.014514} {\bibfield
  {journal} {\bibinfo  {journal} {Phys. Rev. D}\ }\textbf {\bibinfo {volume}
  {108}},\ \bibinfo {pages} {014514} (\bibinfo {year} {2023})}\BibitemShut
  {NoStop}%
\bibitem [{\citenamefont {Yang}\ \emph {et~al.}(2023)\citenamefont {Yang},
  \citenamefont {Zhang}, \citenamefont {Liao}, \citenamefont {Tu},\ and\
  \citenamefont {Wang}}]{Yang2023Mar}%
  \BibitemOpen
  \bibfield  {author} {\bibinfo {author} {\bibfnamefont {Q.}~\bibnamefont
  {Yang}}, \bibinfo {author} {\bibfnamefont {X.-Y.}\ \bibnamefont {Zhang}},
  \bibinfo {author} {\bibfnamefont {H.-J.}\ \bibnamefont {Liao}}, \bibinfo
  {author} {\bibfnamefont {H.-H.}\ \bibnamefont {Tu}},\ and\ \bibinfo {author}
  {\bibfnamefont {L.}~\bibnamefont {Wang}},\ }\bibfield  {title} {\bibinfo
  {title} {{Projected $d$-wave superconducting state: A fermionic projected
  entangled pair state study}},\ }\href
  {https://doi.org/10.1103/PhysRevB.107.125128} {\bibfield  {journal} {\bibinfo
   {journal} {Phys. Rev. B}\ }\textbf {\bibinfo {volume} {107}},\ \bibinfo
  {pages} {125128} (\bibinfo {year} {2023})}\BibitemShut {NoStop}%
\bibitem [{\citenamefont {Li}\ \emph {et~al.}(2023)\citenamefont {Li},
  \citenamefont {von Delft},\ and\ \citenamefont {Tu}}]{Li2023Feb}%
  \BibitemOpen
  \bibfield  {author} {\bibinfo {author} {\bibfnamefont {J.-W.}\ \bibnamefont
  {Li}}, \bibinfo {author} {\bibfnamefont {J.}~\bibnamefont {von Delft}},\ and\
  \bibinfo {author} {\bibfnamefont {H.-H.}\ \bibnamefont {Tu}},\ }\bibfield
  {title} {\bibinfo {title} {{U(1)-symmetric Gaussian fermionic projected
  entangled paired states and their Gutzwiller projection}},\ }\href
  {https://doi.org/10.1103/PhysRevB.107.085148} {\bibfield  {journal} {\bibinfo
   {journal} {Phys. Rev. B}\ }\textbf {\bibinfo {volume} {107}},\ \bibinfo
  {pages} {085148} (\bibinfo {year} {2023})}\BibitemShut {NoStop}%
\bibitem [{\citenamefont {Vieijra}\ \emph {et~al.}(2021)\citenamefont
  {Vieijra}, \citenamefont {Haegeman}, \citenamefont {Verstraete},\ and\
  \citenamefont {Vanderstraeten}}]{Vieijra2021Dec}%
  \BibitemOpen
  \bibfield  {author} {\bibinfo {author} {\bibfnamefont {T.}~\bibnamefont
  {Vieijra}}, \bibinfo {author} {\bibfnamefont {J.}~\bibnamefont {Haegeman}},
  \bibinfo {author} {\bibfnamefont {F.}~\bibnamefont {Verstraete}},\ and\
  \bibinfo {author} {\bibfnamefont {L.}~\bibnamefont {Vanderstraeten}},\
  }\bibfield  {title} {\bibinfo {title} {{Direct sampling of projected
  entangled-pair states}},\ }\href
  {https://doi.org/10.1103/PhysRevB.104.235141} {\bibfield  {journal} {\bibinfo
   {journal} {Phys. Rev. B}\ }\textbf {\bibinfo {volume} {104}},\ \bibinfo
  {pages} {235141} (\bibinfo {year} {2021})}\BibitemShut {NoStop}%
\bibitem [{\citenamefont {Hasik}\ \emph {et~al.}(2022)\citenamefont {Hasik},
  \citenamefont {Van~Damme}, \citenamefont {Poilblanc},\ and\ \citenamefont
  {Vanderstraeten}}]{Hasik2022Oct}%
  \BibitemOpen
  \bibfield  {author} {\bibinfo {author} {\bibfnamefont {J.}~\bibnamefont
  {Hasik}}, \bibinfo {author} {\bibfnamefont {M.}~\bibnamefont {Van~Damme}},
  \bibinfo {author} {\bibfnamefont {D.}~\bibnamefont {Poilblanc}},\ and\
  \bibinfo {author} {\bibfnamefont {L.}~\bibnamefont {Vanderstraeten}},\
  }\bibfield  {title} {\bibinfo {title} {{Simulating Chiral Spin Liquids with
  Projected Entangled-Pair States}},\ }\href
  {https://doi.org/10.1103/PhysRevLett.129.177201} {\bibfield  {journal}
  {\bibinfo  {journal} {Phys. Rev. Lett.}\ }\textbf {\bibinfo {volume} {129}},\
  \bibinfo {pages} {177201} (\bibinfo {year} {2022})}\BibitemShut {NoStop}%
\bibitem [{\citenamefont {Sinha}\ \emph {et~al.}(2024)\citenamefont {Sinha},
  \citenamefont {Rams},\ and\ \citenamefont {Dziarmaga}}]{Sinha2024Jan}%
  \BibitemOpen
  \bibfield  {author} {\bibinfo {author} {\bibfnamefont {A.}~\bibnamefont
  {Sinha}}, \bibinfo {author} {\bibfnamefont {M.~M.}\ \bibnamefont {Rams}},\
  and\ \bibinfo {author} {\bibfnamefont {J.}~\bibnamefont {Dziarmaga}},\
  }\bibfield  {title} {\bibinfo {title} {{Efficient representation of minimally
  entangled typical thermal states in two dimensions via projected entangled
  pair states}},\ }\href {https://doi.org/10.1103/PhysRevB.109.045136}
  {\bibfield  {journal} {\bibinfo  {journal} {Phys. Rev. B}\ }\textbf {\bibinfo
  {volume} {109}},\ \bibinfo {pages} {045136} (\bibinfo {year}
  {2024})}\BibitemShut {NoStop}%
\bibitem [{\citenamefont {Lubasch}\ \emph
  {et~al.}(2014{\natexlab{a}})\citenamefont {Lubasch}, \citenamefont {Cirac},\
  and\ \citenamefont {Ba{\ifmmode\tilde{n}\else\~{n}\fi}uls}}]{Lubasch2014Aug}%
  \BibitemOpen
  \bibfield  {author} {\bibinfo {author} {\bibfnamefont {M.}~\bibnamefont
  {Lubasch}}, \bibinfo {author} {\bibfnamefont {J.~I.}\ \bibnamefont {Cirac}},\
  and\ \bibinfo {author} {\bibfnamefont {M.-C.}\ \bibnamefont
  {Ba{\ifmmode\tilde{n}\else\~{n}\fi}uls}},\ }\bibfield  {title} {\bibinfo
  {title} {{Algorithms for finite projected entangled pair states}},\ }\href
  {https://doi.org/10.1103/PhysRevB.90.064425} {\bibfield  {journal} {\bibinfo
  {journal} {Phys. Rev. B}\ }\textbf {\bibinfo {volume} {90}},\ \bibinfo
  {pages} {064425} (\bibinfo {year} {2014}{\natexlab{a}})}\BibitemShut
  {NoStop}%
\bibitem [{\citenamefont {Lubasch}\ \emph
  {et~al.}(2014{\natexlab{b}})\citenamefont {Lubasch}, \citenamefont {Cirac},\
  and\ \citenamefont {Ba{\ifmmode\tilde{n}\else\~{n}\fi}uls}}]{Lubasch2014Mar}%
  \BibitemOpen
  \bibfield  {author} {\bibinfo {author} {\bibfnamefont {M.}~\bibnamefont
  {Lubasch}}, \bibinfo {author} {\bibfnamefont {J.~I.}\ \bibnamefont {Cirac}},\
  and\ \bibinfo {author} {\bibfnamefont {M.-C.}\ \bibnamefont
  {Ba{\ifmmode\tilde{n}\else\~{n}\fi}uls}},\ }\bibfield  {title} {\bibinfo
  {title} {{Unifying projected entangled pair state contractions}},\ }\href
  {https://doi.org/10.1088/1367-2630/16/3/033014} {\bibfield  {journal}
  {\bibinfo  {journal} {New J. Phys.}\ }\textbf {\bibinfo {volume} {16}},\
  \bibinfo {pages} {033014} (\bibinfo {year} {2014}{\natexlab{b}})}\BibitemShut
  {NoStop}%
\bibitem [{\citenamefont {Scheb}\ and\ \citenamefont
  {Noack}(2023)}]{Scheb2023Apr}%
  \BibitemOpen
  \bibfield  {author} {\bibinfo {author} {\bibfnamefont {M.}~\bibnamefont
  {Scheb}}\ and\ \bibinfo {author} {\bibfnamefont {R.~M.}\ \bibnamefont
  {Noack}},\ }\bibfield  {title} {\bibinfo {title} {{Finite projected entangled
  pair states for the Hubbard model}},\ }\href
  {https://doi.org/10.1103/PhysRevB.107.165112} {\bibfield  {journal} {\bibinfo
   {journal} {Phys. Rev. B}\ }\textbf {\bibinfo {volume} {107}},\ \bibinfo
  {pages} {165112} (\bibinfo {year} {2023})}\BibitemShut {NoStop}%
\bibitem [{\citenamefont {Gleis}\ \emph {et~al.}(2023)\citenamefont {Gleis},
  \citenamefont {Li},\ and\ \citenamefont {von Delft}}]{Gleis2023Jun}%
  \BibitemOpen
  \bibfield  {author} {\bibinfo {author} {\bibfnamefont {A.}~\bibnamefont
  {Gleis}}, \bibinfo {author} {\bibfnamefont {J.-W.}\ \bibnamefont {Li}},\ and\
  \bibinfo {author} {\bibfnamefont {J.}~\bibnamefont {von Delft}},\ }\bibfield
  {title} {\bibinfo {title} {{Controlled Bond Expansion for Density Matrix
  Renormalization Group Ground State Search at Single-Site Costs}},\ }\href
  {https://doi.org/10.1103/PhysRevLett.130.246402} {\bibfield  {journal}
  {\bibinfo  {journal} {Phys. Rev. Lett.}\ }\textbf {\bibinfo {volume} {130}},\
  \bibinfo {pages} {246402} (\bibinfo {year} {2023})}\BibitemShut {NoStop}%
\bibitem [{\citenamefont {Rokhlin}\ \emph {et~al.}(2009)\citenamefont
  {Rokhlin}, \citenamefont {Szlam},\ and\ \citenamefont
  {Tygert}}]{Rokhlin2009Aug}%
  \BibitemOpen
  \bibfield  {author} {\bibinfo {author} {\bibfnamefont {V.}~\bibnamefont
  {Rokhlin}}, \bibinfo {author} {\bibfnamefont {A.}~\bibnamefont {Szlam}},\
  and\ \bibinfo {author} {\bibfnamefont {M.}~\bibnamefont {Tygert}},\
  }\bibfield  {title} {\bibinfo {title} {{A Randomized Algorithm for Principal
  Component Analysis}},\ }\href
  {https://epubs.siam.org/doi/abs/10.1137/080736417} {\bibfield  {journal}
  {\bibinfo  {journal} {SIAM J. Matrix Anal. Appl.}\ } (\bibinfo {year}
  {2009})}\BibitemShut {NoStop}%
\bibitem [{\citenamefont {Halko}\ \emph {et~al.}(2011)\citenamefont {Halko},
  \citenamefont {Martinsson},\ and\ \citenamefont {Tropp}}]{Halko2011May}%
  \BibitemOpen
  \bibfield  {author} {\bibinfo {author} {\bibfnamefont {N.}~\bibnamefont
  {Halko}}, \bibinfo {author} {\bibfnamefont {P.~G.}\ \bibnamefont
  {Martinsson}},\ and\ \bibinfo {author} {\bibfnamefont {J.~A.}\ \bibnamefont
  {Tropp}},\ }\bibfield  {title} {\bibinfo {title} {{Finding Structure with
  Randomness: Probabilistic Algorithms for Constructing Approximate Matrix
  Decompositions}},\ }\href {https://epubs.siam.org/doi/abs/10.1137/090771806}
  {\bibfield  {journal} {\bibinfo  {journal} {SIAM Rev.}\ } (\bibinfo {year}
  {2011})}\BibitemShut {NoStop}%
\bibitem [{\citenamefont {Gleis}\ \emph {et~al.}(2022)\citenamefont {Gleis},
  \citenamefont {Li},\ and\ \citenamefont {von Delft}}]{Gleis2022Nov}%
  \BibitemOpen
  \bibfield  {author} {\bibinfo {author} {\bibfnamefont {A.}~\bibnamefont
  {Gleis}}, \bibinfo {author} {\bibfnamefont {J.-W.}\ \bibnamefont {Li}},\ and\
  \bibinfo {author} {\bibfnamefont {J.}~\bibnamefont {von Delft}},\ }\bibfield
  {title} {\bibinfo {title} {{Projector formalism for kept and discarded spaces
  of matrix product states}},\ }\href
  {https://doi.org/10.1103/PhysRevB.106.195138} {\bibfield  {journal} {\bibinfo
   {journal} {Phys. Rev. B}\ }\textbf {\bibinfo {volume} {106}},\ \bibinfo
  {pages} {195138} (\bibinfo {year} {2022})}\BibitemShut {NoStop}%
\bibitem [{\citenamefont {McCulloch}\ and\ \citenamefont
  {Osborne}(2024)}]{McCulloch2024Mar}%
  \BibitemOpen
  \bibfield  {author} {\bibinfo {author} {\bibfnamefont {I.~P.}\ \bibnamefont
  {McCulloch}}\ and\ \bibinfo {author} {\bibfnamefont {J.~J.}\ \bibnamefont
  {Osborne}},\ }\bibfield  {title} {\bibinfo {title} {{Comment on "Controlled
  Bond Expansion for Density Matrix Renormalization Group Ground State Search
  at Single-Site Costs" (Extended Version)}},\ }\bibfield  {journal} {\bibinfo
  {journal} {arXiv}\ }\href {https://doi.org/10.48550/arXiv.2403.00562}
  {10.48550/arXiv.2403.00562} (\bibinfo {year} {2024}),\ \Eprint
  {https://arxiv.org/abs/2403.00562} {2403.00562} \BibitemShut {NoStop}%
\bibitem [{\citenamefont {Evenbly}(2018)}]{Evenbly2018Aug}%
  \BibitemOpen
  \bibfield  {author} {\bibinfo {author} {\bibfnamefont {G.}~\bibnamefont
  {Evenbly}},\ }\bibfield  {title} {\bibinfo {title} {{Gauge fixing, canonical
  forms, and optimal truncations in tensor networks with closed loops}},\
  }\href {https://doi.org/10.1103/PhysRevB.98.085155} {\bibfield  {journal}
  {\bibinfo  {journal} {Phys. Rev. B}\ }\textbf {\bibinfo {volume} {98}},\
  \bibinfo {pages} {085155} (\bibinfo {year} {2018})}\BibitemShut {NoStop}%
\bibitem [{\citenamefont {Zheng}\ \emph {et~al.}(2017)\citenamefont {Zheng},
  \citenamefont {Chung}, \citenamefont {Corboz}, \citenamefont {Ehlers},
  \citenamefont {Qin}, \citenamefont {Noack}, \citenamefont {Shi},
  \citenamefont {White}, \citenamefont {Zhang},\ and\ \citenamefont
  {Chan}}]{Zheng2017Dec}%
  \BibitemOpen
  \bibfield  {author} {\bibinfo {author} {\bibfnamefont {B.-X.}\ \bibnamefont
  {Zheng}}, \bibinfo {author} {\bibfnamefont {C.-M.}\ \bibnamefont {Chung}},
  \bibinfo {author} {\bibfnamefont {P.}~\bibnamefont {Corboz}}, \bibinfo
  {author} {\bibfnamefont {G.}~\bibnamefont {Ehlers}}, \bibinfo {author}
  {\bibfnamefont {M.-P.}\ \bibnamefont {Qin}}, \bibinfo {author} {\bibfnamefont
  {R.~M.}\ \bibnamefont {Noack}}, \bibinfo {author} {\bibfnamefont
  {H.}~\bibnamefont {Shi}}, \bibinfo {author} {\bibfnamefont {S.~R.}\
  \bibnamefont {White}}, \bibinfo {author} {\bibfnamefont {S.}~\bibnamefont
  {Zhang}},\ and\ \bibinfo {author} {\bibfnamefont {G.~K.-L.}\ \bibnamefont
  {Chan}},\ }\bibfield  {title} {\bibinfo {title} {{Stripe order in the
  underdoped region of the two-dimensional Hubbard model}},\ }\href
  {https://doi.org/10.1126/science.aam7127} {\bibfield  {journal} {\bibinfo
  {journal} {Science}\ }\textbf {\bibinfo {volume} {358}},\ \bibinfo {pages}
  {1155} (\bibinfo {year} {2017})}\BibitemShut {NoStop}%
\bibitem [{\citenamefont {Lin}\ \emph {et~al.}(2024)\citenamefont {Lin},
  \citenamefont {Guo}, \citenamefont {He}, \citenamefont {Xie},\ and\
  \citenamefont {Lu}}]{Lin2024Jun}%
  \BibitemOpen
  \bibfield  {author} {\bibinfo {author} {\bibfnamefont {H.-Y.}\ \bibnamefont
  {Lin}}, \bibinfo {author} {\bibfnamefont {Y.}~\bibnamefont {Guo}}, \bibinfo
  {author} {\bibfnamefont {R.-Q.}\ \bibnamefont {He}}, \bibinfo {author}
  {\bibfnamefont {Z.~Y.}\ \bibnamefont {Xie}},\ and\ \bibinfo {author}
  {\bibfnamefont {Z.-Y.}\ \bibnamefont {Lu}},\ }\bibfield  {title} {\bibinfo
  {title} {{Green's function Monte Carlo combined with projected entangled pair
  state approach to the frustrated ${J}_{1}\text{\ensuremath{-}}{J}_{2}$
  Heisenberg model}},\ }\href {https://doi.org/10.1103/PhysRevB.109.235133}
  {\bibfield  {journal} {\bibinfo  {journal} {Phys. Rev. B}\ }\textbf {\bibinfo
  {volume} {109}},\ \bibinfo {pages} {235133} (\bibinfo {year}
  {2024})}\BibitemShut {NoStop}%
\bibitem [{\citenamefont {Liu}\ \emph {et~al.}(2025)\citenamefont {Liu},
  \citenamefont {Zhai}, \citenamefont {Peng}, \citenamefont {Gu},\ and\
  \citenamefont {Chan}}]{Liu2025Jun}%
  \BibitemOpen
  \bibfield  {author} {\bibinfo {author} {\bibfnamefont {W.-Y.}\ \bibnamefont
  {Liu}}, \bibinfo {author} {\bibfnamefont {H.}~\bibnamefont {Zhai}}, \bibinfo
  {author} {\bibfnamefont {R.}~\bibnamefont {Peng}}, \bibinfo {author}
  {\bibfnamefont {Z.-C.}\ \bibnamefont {Gu}},\ and\ \bibinfo {author}
  {\bibfnamefont {G.~K.-L.}\ \bibnamefont {Chan}},\ }\bibfield  {title}
  {\bibinfo {title} {{Accurate Simulation of the Hubbard Model with Finite
  Fermionic Projected Entangled Pair States}},\ }\href
  {https://doi.org/10.1103/r4q9-4yvj} {\bibfield  {journal} {\bibinfo
  {journal} {Phys. Rev. Lett.}\ }\textbf {\bibinfo {volume} {134}},\ \bibinfo
  {pages} {256502} (\bibinfo {year} {2025})}\BibitemShut {NoStop}%
\bibitem [{\citenamefont {Tindall}\ and\ \citenamefont
  {Fishman}(2023)}]{Tindall2023Dec}%
  \BibitemOpen
  \bibfield  {author} {\bibinfo {author} {\bibfnamefont {J.}~\bibnamefont
  {Tindall}}\ and\ \bibinfo {author} {\bibfnamefont {M.}~\bibnamefont
  {Fishman}},\ }\bibfield  {title} {\bibinfo {title} {{Gauging tensor networks
  with belief propagation}},\ }\href
  {https://doi.org/10.21468/SciPostPhys.15.6.222} {\bibfield  {journal}
  {\bibinfo  {journal} {SciPost Phys.}\ }\textbf {\bibinfo {volume} {15}},\
  \bibinfo {pages} {222} (\bibinfo {year} {2023})}\BibitemShut {NoStop}%
\bibitem [{\citenamefont {Evenbly}\ \emph {et~al.}(2024)\citenamefont
  {Evenbly}, \citenamefont {Pancotti}, \citenamefont {Milsted}, \citenamefont
  {Gray},\ and\ \citenamefont {Chan}}]{Evenbly2024Sep}%
  \BibitemOpen
  \bibfield  {author} {\bibinfo {author} {\bibfnamefont {G.}~\bibnamefont
  {Evenbly}}, \bibinfo {author} {\bibfnamefont {N.}~\bibnamefont {Pancotti}},
  \bibinfo {author} {\bibfnamefont {A.}~\bibnamefont {Milsted}}, \bibinfo
  {author} {\bibfnamefont {J.}~\bibnamefont {Gray}},\ and\ \bibinfo {author}
  {\bibfnamefont {G.~K.-L.}\ \bibnamefont {Chan}},\ }\bibfield  {title}
  {\bibinfo {title} {{Loop Series Expansions for Tensor Networks}},\ }\bibfield
   {journal} {\bibinfo  {journal} {arXiv}\ }\href
  {https://doi.org/10.48550/arXiv.2409.03108} {10.48550/arXiv.2409.03108}
  (\bibinfo {year} {2024}),\ \Eprint {https://arxiv.org/abs/2409.03108}
  {2409.03108} \BibitemShut {NoStop}%
\end{thebibliography}%

\end{document}